\documentclass[fleqn,usenatbib]{mnras}

\usepackage{newtxtext,newtxmath}

\usepackage[T1]{fontenc}
\usepackage{ae,aecompl}
\usepackage{soul}

\usepackage{graphicx}	
\usepackage{amsmath}	
\usepackage{amssymb}	
\usepackage{multicol}   
\usepackage{pdflscape}	
\usepackage{rotating}   
\usepackage{caption}    
\usepackage{longtable}  
\usepackage[normalem]{ulem} 
\usepackage{color}
\usepackage{wasysym}
\usepackage{wedn}
\usepackage{miama}
\usepackage{calligra} 
\usepackage[Symbol]{upgreek}
\DeclareTextFontCommand{\textmyfont}{\calligra}
\usepackage[dvipsnames]{xcolor}

\usepackage{etoolbox}
\makeatletter
\patchcmd\@combinedblfloats{\box\@outputbox}{\unvbox\@outputbox}{}{%
   \errmessage{\noexpand\@combinedblfloats could not be patched}%
}%
 \makeatother
 \usepackage[dvipsnames]{xcolor}
 \definecolor{black}{rgb}{34,156,71}
  \newcommand{\ARTISTtitle}{{\fontsize{30}{104}{\fontfamily{fmm}\selectfont {A}}}{\sc rtist}}  
  \newcommand{\ARTIST}{{\huge{\fontfamily{fmm}\selectfont {A}}}{\sc rtist}} 

\title[Fast RT-based simulations of the EoR]{\ARTISTtitle{}: Fast radiative transfer for large-scale simulations of the epoch of reionisation}

\author[Molaro et al.]{Margherita Molaro,$^{1}$\thanks{E-mail: mmolaro@uwc.ac.za}
Romeel Dav\'e,$^{3,1,2}$
Sultan Hassan,$^{4,1}$\thanks{Tombaugh Fellow}
Mario G. Santos,$^{1,5,6}$ \and
Kristian Finlator$^{4,7}$\\\\
$^{1}$ Department of Physics and Astronomy, University of the Western Cape, Bellville, Cape Town, 7535, South Africa\\
$^{2}$ South African Astronomical Observatory, Observatory, Cape Town 7925, South Africa  \\
$^{3}$ Institute for Astronomy, Royal Observatory, Edinburgh EH9 3HJ, UK \\
$^{4}$ New Mexico State University, Las Cruces, NM 88003, USA \\
$^{5}$ South African Radio Astronomy Observatory, Black River Park, 2 Fir Street, Observatory, Cape Town, 7925, South Africa\\
$^{6}$ Instituto de Astrofisica e Ciencias do Espaco, Universidade de Lisboa, OAL, Tapada da Ajuda, PT1349-018 Lisboa, Portugal\\
$^{7}$ Cosmic Dawn Center at the Niels Bohr Institute, University of Copenhagen and DTU-Space, Technical University of Denmark, Elektrovej Bygning 327, \\ 2800 Kongens Lyngby, Denmark
}

\date{Accepted 2019 July 31. Received 2019 July 21; in original form 2018 December 26}

\pubyear{2018}

\begin{document}
\label{firstpage}
\pagerange{\pageref{firstpage}--\pageref{lastpage}}
\maketitle

\begin{abstract}
We introduce the ``Asymmetric Radiative Transfer In Shells Technique'' (\ARTIST{}), a new method for photon propagation on large scales that explicitly conserves photons, propagates photons at the speed of light, approximately accounts for photon directionality, and closely reproduces results of more detailed radiative transfer (RT) methods. Crucially, it is computationally fast enough to evolve the large cosmological volumes required to predict the 21cm power spectrum on scales that will be probed by future experiments targeting the Epoch of Reionisation (EoR). Most semi-numerical models aimed at predicting the EoR 21cm signal {on these scales} make use of an excursion set formalism (ESF) {to model the gas ionisation, which achieves computational viability by making a number of approximations. While \ARTIST{} is still roughly two orders of magnitude slower than ESF, it does allow to model the EoR without the need for such approximations. This is particularly important when considering a wide range of reionisation scenarios for which \ARTIST{} would help limit the assumptions made}. By implementing our RT method within the semi-numerical code {\sc SimFast21}, we show that \ARTIST{} predicts a significantly different evolution for the EoR ionisation field compared to the code's native ESF. In particular, \ARTIST{} predicts {up to a factor of two difference in the power spectra}, depending on the physical parameters assumed.  Its application to large-scale EoR simulations will therefore allow more physically-motivated constraints to be obtained for key EoR parameters. {In particular, it will remove the need for the artificial rescaling of the escape fraction.}

\end{abstract}


\section{Introduction}
\label{intro}
A common challenge faced by astrophysical simulations is having to accurately account for physical processes taking place on a wide range of dynamical scales, while ensuring a feasible computational cost. Radiative transfer (RT) -- a fundamental driver of systems' dynamics in a wide range of astrophysical cases -- remains among the most expensive processes to implement in a computationally self-consistent manner.  As such, many approaches to modelling RT have been developed in astrophysical simulations, including Monte Carlo (MC), long characteristics (i.e. ray tracing), and moment-based methods, with different strengths and weaknesses, balancing accuracy versus computational efficiency \citep{Trac2011}. Often, while these methods can be made optimally accurate with sufficient computational investment, they remain computationally prohibitive in large-scale cosmological simulations that seek to reproduce the evolution of the Universe on at least tens of Mpc scales, while simultaneously ensuring that the injection and propagation of photons on the smallest scales is both accurate and self-consistent.

One particular such case, currently at the forefront of astrophysical research, is the modelling of the last global phase-change in the history of the Universe -- the epoch of reionisation (EoR). The sources of the photons responsible for the reionisation of the intergalactic medium (IGM) between redshift $z \sim 6-20$ are generally believed to be small primeval galaxies, with typical sizes $\leq 1$ kpc \citep{Stark2016,Dayal2018} which began to form at $z\ga 20$. As reionisation proceeds, ionised regions should approach sizes of 100 Mpc before finally overlapping \citep{Loeb2001,Furlanetto2005}. Observations of neutral hydrogen absorption in quasar spectra constrain the redshift by which reionisation is completed to be $z \sim 6$ \citep{Fan2006}.  

Much effort is currently being invested into understanding how exactly the Universe evolved during this phase, from a globally neutral state to an ionised one. A highly promising approach to observationally probing this global phase transition is the large-scale intensity mapping of the hyper-fine, 21cm transition line emitted by neutral hydrogen \citep{Barkana2001}. Due to this emission occurring at a particular rest frequency, radio interferometers are in principle able to reconstruct the morphology of the ionised regions at different redshifts during the EoR. Current and future redshifted 21cm telescopes such as the Low Frequency Array (LOFAR; \citealt{vanHaarlem2013}), the Precision Array for Probing the Epoch of Reionization (PAPER; \citealt{Parsons2012}), the Hydrogen Epoch of Reionization Array (HERA; \citealt{DeBoer2017}), the Murchison Wide field Array (MWA; \citealt{Bowman2013,Tingay2013}), the Giant Metre-wave Radio Telescope (GMRT; \citealt{Paciga2011}), and the Square Kilometre Array (SKA; \citealt{Mellema2015}), will dramatically expand the range and sensitivity with which we will be able to observe the large scale features of 21cm emission in the EoR redshift range ($z\sim 6-10+$).  However, the low frequency of these telescopes naturally leads to large beam sizes, which means that these facilities will typically probe the topology of the neutral gas distribution on fairly large ($\ga$Mpc) scales.

The dynamic range required to model the sources and large-scale 21cm topology of reionisation thus represents a particularly difficult computational challenge.  On one hand, sub-kpc resolution is required to resolve the processes -- atomic cooling, radiative transfer and feedback -- that are crucial to correctly reproduce the population of sources responsible for producing the ionising radiation, {which could reside in halo masses as small as $10^8 \text{M}_{\odot}$ \citep{Iliev2015}}. The minimum mass for efficient star formation has been explored thoroughly from a theoretical perspective by \cite{Noh2014}, while model-dependent observational constraints are given by \cite{Finlator2017}. On the other hand, the scales on which 21cm signal fluctuations become apparent require simulation volumes of $\gg 100$ Mpc in size \citep{Iliev2014}. This implies tracking ionisation and feedback processes across over $\ga 10^5$ orders of magnitude in scale.\\

{Current computational resources are far from being able to meet this challenge directly.  
 Simulations of the EoR must simultaneously account for:
\begin{itemize}
    \item[] i) the evolution of the density field and formation of ionising sources;
    \item[] ii) the propagation of ionising photons through the IGM and consequent ionisation and recombination processes; 
\end{itemize}
Many techniques exist to model these dynamics. With respect to i), these include:
\begin{itemize}
\item semi-numerical simulations of density fields (evolved using a Zeldovich approach \citet{Zeldovich1970}) followed by the identification of collapsed objects through an excursion set formalism (ESF);
\item N-body simulations of density fields, collapsed objects and halos from N-body and/or hydrodynamical simulations.
\end{itemize}
With respect to ii), on the other hand, these include fully radiative transfer methods (either on-the-fly or in post-processing) or ESF approaches to identifying ionised regions.}
{
Notice that ESF can therefore be applied in two contexts in EoR simulations: one concerning the identification of collapsed objects, and one the identification of ionised regions. To avoid confusion, we clarify that all references to ESF methods made from now on refer to the latter. To refer to the former, we will state explicitly that the ESF, in that case, is applied to density fields. }

{Different approaches to i) and ii) can therefore be combined in various ways to obtain a full EoR simulation. As a result, this computational challenge has so far been tackled in the following ways:}

\begin{itemize}
\item {\textbf{N-body, hydrodynamical simulations including self-consistent RT:} These meet the resolution requirements needed to self-consistently model the formation of ionising sources and physically propagate the ionising radiation in the IGM. However, computational constraints to fully achieve these objectives limit the cosmological simulation boxes to sizes of at most tens of Mpc, which is sub-optimal for predicting the 21cm signal fluctuations  \citep{Gnedin2000,Trac2007,Gnedin2014,Pawlik2008,Finlator2009,Finlator2013,Graziani2013,Katz2017}.}
\item {\textbf{N-body, hydrodynamical simulations combined with RT post-processing:} These simulations can probe bigger volumes (up to $\sim 100+$~Mpc), but fail to self-consistently account for processes leading to the formation of ionising sources, and the co-evolution of the source and sink populations  \citep{Razoumov2002,Ciardi2003,Sokasian2001, Mellema2006,McQuinn2007,Semelin2007,Altay2008,Aubert2008, Finlator2009,Thomas2009,Petkova2009,Iliev2014,Bauer2015}.}
\item {{N-body, hydrodynamical simulations combined with ESF:}} {N-body simulations of the density field and collapsed halos coupled with an ESF approach to identify ionised regions \citep{Zahn2007,Hutter2018}}
\item {\textbf{{ESF-based halo-identification with ESF-based ionisation schemes:}} These combine a quasi-linear evolution of the density field and an ESF approach to collapsed-objects-identification, with parametrised relations linking dark matter (DM) halos to UV photons emission, and recombination rates to the hydrogen overdensity.
The ionising photons in these semi-numerical simulations are then \textit{not} propagated using a RT approach, but rather their contribution to the ionisation of the IGM is approximated using the ESF method \citep{Press1974,Bond1991}, which we recap below \citep[see also][]{Mesinger2007,Zahn2007,Geil2008,Alvarez2009,Choudhury2009,Santos2010}.}

\end{itemize}
{
The latter semi-numerical approaches are currently the only ones that can predict the ionisation evolution of the IGM on the large scales that will be probed by future radio experiments. These can approximately account for the gas and photon dynamics at the smallest scales through the use of the parametrised relations to connect ionising photon production to halo growth. Hence at present, ESF models are usually the approach of choice for EoR 21cm forecasting.}\\

The use of ESF in these models, however, incurs several significant limitations. One of the most crucial ones is the intrinsic difficulty for ESF simulations to accurately conserve the number of ionising photons in overlapping regions \citep{McQuinn2005,Zahn2007,Paranjape2014,Paranjape2016,Hassan2017,Hutter2018} with potentially severe errors in predicting the evolution of the neutral fraction. In \cite{Zahn2007}, it was speculated that this discrepancy was overall no more than 20\%, {a value similar to the one found by \cite{Hutter2018}}. \cite{Hassan2017}, however, found that the discrepancy could be significantly higher at particular times, most notably during the time when the Universe is $\sim 50\%$ neutral, which is a key target phase for 21cm experiments. \\
{The relation between the assumed ionising sources population and the simulated EoR evolution is  compromised by such issues of photon non-conservation.} A previously unrecognised consequence of this issue was recently presented by \cite{Choudhury2018}, who found that a few-percent error in the photon conservation can lead to a strong resolution dependence in the HI bias, which could lead to a deviation from the converged value by as much as 20-25\% at a resolution of $\Delta x  = 5-10 ~c\text{Mpc/h}$ for a photon conservation error as low as 3-4\% . Given advancing multi-wavelength EoR observations, it is unclear that the assumptions intrinsic to ESF are adequate to accurately connect 21cm  observables to the topology and the underlying source population in the EoR, which is a key goal of EoR 21cm modeling.  This highlights the need to develop more accurate RT methods to study the EoR, while maintaining computational tractability {of larger scales}.

Several recent works have attempted to address the limitations of ESF methods in the context of EoR simulations in different ways, {for example by considering ESFs based single-cell rather than whole-sphere flagging, first introduced by \citet{Zahn2007}, and later expended by \citet{Mesinger2011} and more recently \cite{Hutter2018}.} \cite{Choudhury2018} suggested that a new method for post-processing overlapping ionised regions could help overcome issues of photon conservation. 
 
{
In this work, we take a different approach and introduce a new RT method that - while not able to fully achieve the computational performance of ESF-based simulations - provides a relatively fast RT method which removes the need for such assumptions and ad-hoc corrections altogether. }

Out new RT method approximates a full Monte Carlo approach, with an implementation that is computationally feasible across the required dynamic range. For reasons that will become evident, we refer to this approach as the ``Asymmetric Radiative Transfer In Shells Technique'' (\ARTIST). Because its degree of approximation can be freely constrained (balanced against computational cost), this method is highly flexible and can be adjusted to the particular requirements of individual use cases. The explicit nature of the physical assumptions made by this technique -- which crucially distinguishes \ARTIST{} from ESF methods -- further allows one to conduct numerical convergence studies in order to directly estimate the inaccuracies introduced by these approximations. 

In the first part of the paper, we discuss the principles and implementation of \ARTIST{} (section \ref{algorithm}), and test its accuracy by comparing it with other, more accurate RT methods available (section \ref{compRT}). In the second part of the paper, we consider its application to the test case of semi-numerical simulations of the EoR, and carry out comparisons against the {whole-sphere} ESF approach to quantify the differences between the two methods (section \ref{esf}), and provide computational benchmarks for \ARTIST{} in section \ref{performance}.  We  recap and summarise our findings in section \ref{conclusion}.

\section{A new RT method: \ARTIST}
\label{algorithm}

   \begin{figure*}
   \centering
   \includegraphics[trim = 200 0 0 50, height=5cm]{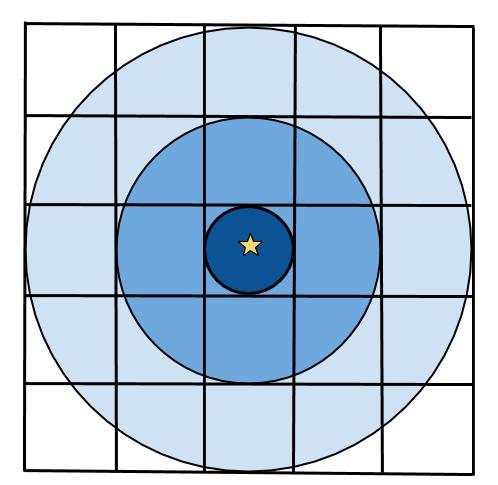} 
   \includegraphics[trim =180 0 80 0, clip, height=5.5cm]{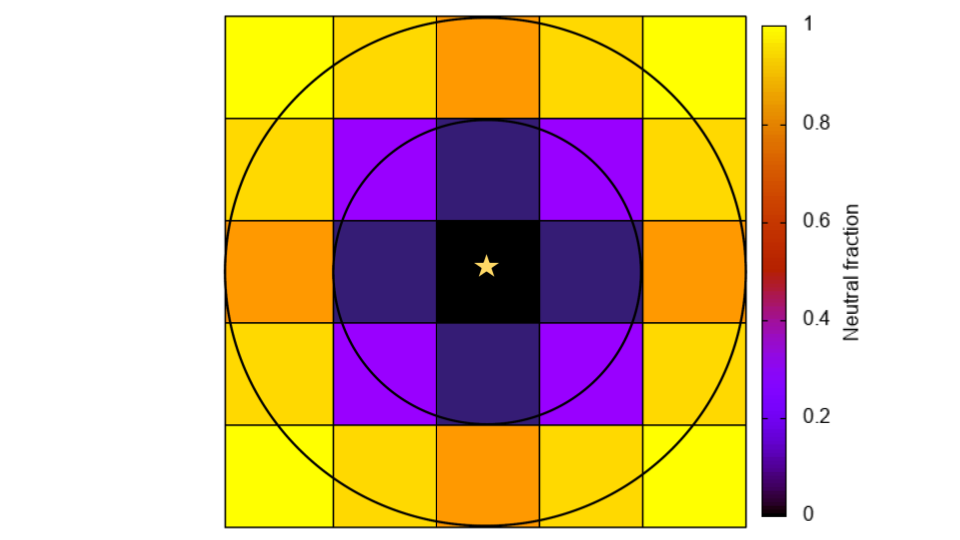}
   \includegraphics[trim = 0 0 200 50, height=5cm]{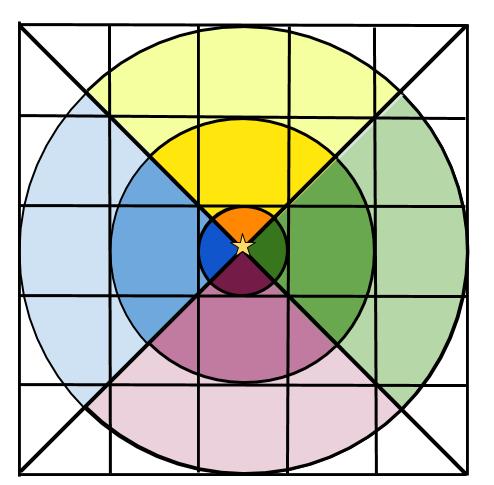}
      \caption{\textit{Left diagram}: Diagram illustrating the propagation of shells through a grid, each with a thickness corresponding to the cell size. \textit{Middle diagram}: partial ionisation of cells in the grid due to the propagation of photons with the \ARTIST{} method. \textit{Right diagram}: Diagram illustrating the resolution of the propagating shells into different solid angles or ``pyramids'', to allow for a better resolution of the propagating shells across inhomogeneous density distributions. To improve the degree of accuracy of the propagation, the number of directions considered ($D$) can be arbitrarily increased (at a computational cost).
              }
              
         \label{cartoon}
   \end{figure*}

We start by briefly reviewing the basic theory of the physical processes taking place during the EoR in section \ref{review}. We then discuss the principles of our method by first considering the case of a single source (section \ref{single_source}) in a uniform medium (subsection \ref{single_source_hom}), then that of a single source in an inhomogeneous medium (subsection \ref{asymmetric_prop}), and finally generalise this to multiple sources (section \ref{multi_source}). Finally, we discuss the propagation of photons in a more diffuse background, once many ionised regions overlap (section \ref{background}). A summary of the main features of the algorithm is given in section \ref{summary}. All parameters defined in this section are summarised in Table \ref{tab:summary} for convenience.

\subsection{Photon propagation in the EoR}
\label{review}
The study of the Epoch of Reionisation is largely the study of the evolution of the IGM's neutral hydrogen content as this is illuminated by the first luminous sources. This is due to the HI's 21cm line emission being by far the most abundant source of information we can observe from this phase of the Universe's history.

Reproducing the evolution of HI over this period is therefore a primary objective of any simulation aiming to study the EoR. As the UV photons emitted by the first galaxies propagate through the IGM, two competing processes affect the hydrogen's ionisation state, and the subsequent propagation of photons through the same medium: photo-ionisation and recombination. The relative contribution of these two processes given local conditions ultimately determines the morphology and dynamics of the hydrogen's evolution from a neutral to an ionised phase. Reproducing it is therefore the primary objective of all simulations seeking to study this epoch. 
In the case of a gridded simulation, this translates into being able to self-consistently account for the following quantities within each grid-cell, at every time-step:
 \begin{itemize}
     \item The number of photoionising photons, $\gamma$;
     \item The ionisation state of the cell, or the fraction of neutral to  total hydrogen atoms $x_{\text{HI}}$; and
     \item The number of recombinations and photoionisations taking place in the cell, given the above two quantities.
     \end{itemize}

Accounting for the presence of photons in each cell at every time-step is by far the most challenging and computationally demanding aspect of such simulations. This is due to the fact that all photoionisation and recombination processes along the line of sight of each photon must be taken into account self-consistently as the photons propagate. Furthermore, since considering each individual photon emitted by every source at every time-step of the simulation is beyond the current limits of computational capability, approximations have to be introduced to capture the propagation of individual photons throughout the 3D volume surrounding the sources. Such approximations must be carried out in a way that, while being computationally viable, don't compromise on the accuracy of the physical processes being simulated.

In the following sections, we outline how our method of photon propagation in gridded cosmological volumes, \ARTIST, aims to accomplish this. 

\begin{table*}
\caption{Parameter table}
\label{tab:summary}
\begin{tabular}{c l}
\hline
Parameter & Description \\
\hline

$V_{s,d}$ &  Volume of shell $s$ in direction $d$, referred to as shell section ($s,d$)\\
$v_i$ & Volume of cell $i$ \\
$v_{s,d,i}$ & Volume of shell section ($s,d$) inside cell $i$ \\
$f_{\text{shell},s,d,i} = v_{s,d,i}/V_{s,d}$ & Fraction of volume of shell section ($s,d$) found inside cell $i$ \\
$f_{\text{cell},s,d,i} = v_{s,d,i}/v_i $ & Fraction of cell $i$'s volume occupied by shell section ($s,d$) \\
$\gamma_{s,d,i}$ & Photons in shell section ($s,d$), illuminating cell $i$ \\
$N_{\text{H},i}$ & Hydrogen atoms (neutral and ionised) in cell $i$ \\ 
$x_{\text{HI},i}$ & Neutral HI fraction in cell $i$ \\
$x_{\text{HI},s,d,i}$ & Neutral HI fraction in fraction of cell $i$'s volume illuminated by $\gamma_{s,d,i}$ photons\\
$\gamma_{l,s,d,i}$ & Photons originally emitted by source $l$, in shell section ($s,d$), illuminating cell $i$ \\
$\epsilon$ & Cut-off to cumulative contribution of photon packages inside cells to determine background transition. \\
\hline
\end{tabular}
\end{table*}

\subsection{Single source case }
\label{single_source}
\subsubsection{Spherical photon propagation}
\label{single_source_hom}
Photons emitted isotropically in a time-interval $\text{d}t$ illuminate a shell of thickness $\text{d}t \times c$ and inner radius $\text{d}t \times (T-1) \times c$ around the source (where $T$ is the number of time-steps from the time the photon was emitted, and $c$ is the speed of light). In our method, we consider such propagating shells for a thickness equal to the cell size d$x$ in our simulation grid, such that:
\begin{equation}
\label{timestep}
\text{d}t= c/\text{d}x.
\end{equation}

Because the shell's volume is spherical, it illuminates cells in the squared grid by different amounts, depending on their relative position to the source. The shell's volume $V_s$ is therefore split across a number of cells, with certain cells containing more of it than others, as illustrated (in 2-D) in Fig. \ref{cartoon}. Given the volume $v_{s,i}$ of shell $s$ found inside cell $i$, the fraction of the volume of shell $s$ that cell $i$ contains is:
\begin{equation}
    f_{\text{shell},s,i} =  v_{s,i}/V_{s}
\end{equation}

\noindent where $i$ runs over all the cells in the simulation volume. By definition, therefore:
\begin{equation}
\label{f_shell}
\sum_i f_{\text{shell},s,i} = 1
\end{equation}

If we assume that photons in shell $s$ are homogeneously distributed within it, then $f_{\text{shell},s,i}$ also determines the fraction of $s$-shell photons that each cell contains. \ARTIST{} makes use of this information to propagate the photons across the grid in subsequent time-steps, as illustrated below.

Assume that a source of ionising radiation, located in cell $i=0$, emits $R_{\text{ion}}$ photons per second. In a time-step $\text{d}t$, a total number of photons, $\gamma = R_{\text{ion}} \times \text{d}t$, will be emitted in a shell of inner radius 0, which we label $s = 0$. These photons will then ionise the neutral hydrogen in the cell. 

Now, assume that a number of photons $\gamma' = \gamma'_{s = 0}$ is left over following absorption and recombination processes in the cell after illumination by shell $s = 0$ photons (the method to calculate leftover photons in a cell will be discussed in section \ref{leftoverincell}). We redistribute these photons across the cells that will be illuminated by the next shell $s = 1$ using pre-calculated $f_{\text{shell},s=1,i}$ values, such that: 
\begin{equation}
\label{redistr}
\gamma_{s=1,i} = \gamma'_{s=0} \times f_{\text{shell},s=1,i}
\end{equation}
where $\gamma_{s=1,i}$ is the number of photons in shell $s = 1$ that will be assigned to cell $i$.

At the next time-step, the total number of photons leftover from illumination by shell $s=1$ will then be:
\begin{equation}
\label{leftphot}
\gamma'_{s=1} = \sum_i \gamma'_{s=1,i}
\end{equation}
where $i$ runs over all the cells which were illuminated by $s=1$. 
These will then themselves be redistributed across $s=2$ using the same method of relative volume contribution explained above, and so on for all subsequent shells. Eqns. \ref{redistr} and \ref{leftphot} can therefore be generalised as:
\begin{equation}
\label{gen_redistr}
\gamma_{s+1,i} = \gamma'_{s} \times f_{\text{shell},s+1,i}
\end{equation}
and
\begin{equation}
\label{gen_collect}
\gamma'_{s} = \sum_i \gamma'_{s,i}
\end{equation}
respectively.

The middle diagram in Fig. \ref{cartoon} illustrates how this results in an ionisation of the squared grid which accurately reflects the propagation of a spherical shell of photons through an homogeneous neutral hydrogen medium. Cells diagonally further away from the source receive fewer photons and at later time, and are therefore less ionised than the others.

\subsubsection{Asymmetric photon propagation}
\label{asymmetric_prop}
In the case of a homogeneous density distribution around the source (one such case will be discussed in section \ref{stromgr}), the spherical averaging of the leftover photons resulting from the sum over cells $i$ in Eqn. \ref{gen_collect} is fully accurate.
However, in the case of an asymmetric density distribution, this introduces certain inaccuracies: some regions around the source may have a high hydrogen density, so that very few photons should be able to cross them. Others might have a low density, so that most of the photons should be able to propagate through them. Adding all leftover photons from a shell together, and then redistributing them in a spherically-averaged way across the next shell, as illustrated above, will hence smear the effect of the inhomogeneous density distribution on the propagation of the photons.

In order to mitigate this, we split the volume of each shell in different directions by ``pyramids'', as illustrated in the third (right) diagram in Fig. \ref{cartoon}. Photons within a given cell will be further identified -- together with their shell number $s$ -- by direction index (or ``pyramid number'') $d$, which identifies the direction the photon is propagating in. We refer to these as shell sections ($s,d$). The propagation of photons can then be carried out for each shell section separately (see again Fig. \ref{cartoon}). Therefore, left over photons will be distinguished both by shell \textit{and} direction. 

Eqn. \ref{gen_redistr} can then be generalised as:
\begin{equation}
\label{gen_redistr_d}
\gamma_{s,d,i} = \gamma'_{s-1,d} \times f_{\text{shell},s,d,i}
\end{equation}
where $f_{\text{shell},s,d,i}$ is defined as: 
\begin{equation}
\label{f_shell_def}
f_{\text{shell},s,d,i} = v_{s,d,i}/V_{s,d}
\end{equation}
and $v_{s,d,i}$ is the volume of shell section ($s,d$) contained in cell $i$, and $V_{s,d}$ is the total volume of shell section ($s,d$).

Eqn. \ref{gen_collect}, on the other hand, is rewritten as:
\begin{equation}
\label{gen_collect_d}
\gamma'_{s,d} = \sum_i \gamma'_{s,d,i}
\end{equation}

By propagating left-over photons independently in each direction $d$, regions located beyond low density ones will be able to receive more leftover photons than regions beyond high density ones. Notice that the accuracy of this method decreases further away from the source, as the number of cells illuminated by later shells increases. 

The accuracy of the method increases as the solid angle of the pyramids decreases, or similarly when the number of pyramids considered increases. In fact, by allowing the solid angle to tend to zero (or allowing the number of possible directions to approach infinity), and then randomly sampling the directions for practical purposes, this method reduces to a Monte Carlo RT approach.

This makes this approach extremely flexible in its accuracy: depending on the computational requirement of the simulation it is applied to, the number of sections considered can be independently constrained to yield a more or less accurate approximation of the radiation transfer.  

In the runs here, we fix the number of directions (or pyramids) to 6, corresponding to the $\pm x,y,z$ axes. Despite the relatively coarse approximation, this is a significant improvement over the ESF approach, which by construction doesn't account for asymmetric density distributions around the source.  In principle this feature of \ARTIST\ enables us to account for self-shielding and shadowing; we will present a more quantitative discussion of this in section \ref{shadowing}.

\subsubsection{Photoionisations and recombinations inside cells}
\label{leftoverincell}
Now that we have explained how \ARTIST{} propagates the leftover photons $\gamma'$ across the grid, here we discuss how $\gamma'$ is calculated.  The basic approach is that each cell produces ionising photons that are added spherically to the directional photons received from other cells, which are then attenuated via recombinations to yield the leftover photons that will be (directionally) propagated to surrounding cells in the next timestep.  Below we describe this more formally.

At each time-step, a cell is assigned a certain number of photons using the method discussed in the previous section. These photons are distinguished by shell number and direction, so that leftover photons can be stored independently for each ($s,d,i$). Each $(s,d)$ packet inside the cell $i$ illuminates a different fraction of the cell's volume. We can define this fraction as:
\begin{equation} 
\label{ffin}
f_{\text{cell}, s,d,i} = v_{s,d,i}/v_i.
\end{equation}
By definition, therefore, 
\begin{equation}
\sum_s \sum_d f_{\text{cell},s,d,i} = 1
\end{equation}
Because photon packets are dynamically allocated (see section \ref{performance}), the splitting of the cell's volume between different shells described by Eqn. \ref{ffin} only takes place if the cell contains any shell-photons at all. 
We stress here that $f_{\text{cell},s,d,i} \neq f_{\text{shell},s,d,i}$ (see parameter summary in Table \ref{tab:summary}).

Because $f_{\text{cell},s,d,i}$ determines the fraction of the cell $i$'s volume that photons $\gamma_{s,d,i}$ are able to illuminate, it also determines (assuming a homogeneous distribution of hydrogen atoms inside the cell) how many hydrogen atoms those photons are able to ionise. The number of neutral atoms left-over in the region illuminated by ($s,d$) at the end of the time-step, $N'_{\text{HI},s,d,i}$, will then be:

\begin{equation}
\begin{split}
\label{leftNHI}
N'_{\text{HI},s,d,i} = & N_{\text{H},i} \times f_{\text{cell},s,d,i} \times x_{\text{HI},s,d,i}   \\ 
& - [\gamma_{s,d,i} - R_{\text{rec}}\text{d}t \times f_{\text{cell},s,d,i} (1-x_{\text{HI},s,d,i})]
\end{split}
\end{equation}

Where the dash refers to values at the end of the time-step; $N_{\text{H},i}$ is the total number of hydrogen atoms in the entire cell (neutral \textit{and} ionised); $x_{\text{HI},s,d,i}$ is the neutral fraction in the region illuminated by $\gamma_{s,d,i}$ photons; and $R_{\text{rec}} \times \text{d}t$ is the total number of recombinations taking place in that cell in that time-step. 

The recombination rates $R_{\text{rec}}$, the number of hydrogen atoms $N_{\text{H}}$, and the photon emission rate $R_{\text{ion}}$, are inputs given to \ARTIST{} by the simulation it is run on. For the test case of its application to {\sc SimFast21}, the $R_{\text{rec}}$,  $R_{\text{ion}}$ and density distribution calculations are discussed in section \ref{recrioN_subsec}. This implies that, although here we apply \ARTIST{} to {\sc SimFast21}'s evolving density field, our method can more generally be used as an on-the-fly RT method for any simulation that can provide $R_{\text{rec}}$,  $R_{\text{ion}}$ and a hydrogen density distribution.

The overall ionisation state in the cell at the end of the time-step will thus be:
\begin{equation}
\label{ioni_fraction} 
x'_i = \frac{ \sum_s \sum_d N'_{\text{HI},s,d,i}}{N_{\text{H},i}}
\end{equation}

It is important to emphasise that the sum over $s$ and $d$ in Eqn. \ref{ioni_fraction} must be carried over the entire volume of the cell,  i.e. even in regions of the cells which haven't yet been reached by photons.

Notice that, although we ultimately output a single ionisation state for each cell, this allows us to keep track of the different ionisation states in each fraction of the cell's volume $f_{\text{cell},s,d,i}$ separately. Therefore the resolution of the simulation with respect to the ionisation distribution is in effect higher than the cell resolution, and optimises the resolution more efficiently than by increasing the number of cells in a Cartesian grid since this would require resolving a curved shell surface with smaller cubical cells.

Note that, from Eqn. \ref{leftNHI}, a negative value of $N'_{\text{HI},s,d,i}$ indicates that cell $i$ contained more photons than neutral hydrogen atoms they could ionise. The number of excess photons from component ($s,d,i$) will therefore be:

\begin{equation}  
\label{photons_left}  
\gamma'_{s,d,i} =
\begin{cases}
|N'_{\text{HI},s,d,i}| & \text{ if } N'_{\text{HI},s,d,i} < 0 \\
0 & \text{ if } N'_{\text{HI},s,d,i} \geq 0 \\
\end{cases}
\end{equation}

These leftover photons will then be propagated across the grid using the method discussed in section \ref{asymmetric_prop}. At the end of the time step, a negative $N’_{\text{HI},s,d,i}$ is reset to zero for the purpose of computing $x_{\text{HI}}$ once its absolute value has been used to propagate the excess photons.

The above sections have described photon propagation in \ARTIST\ from a single source.  A realistic EoR simulation will have multiple sources, eventually resulting in overlapping ionised regions.  
In the next section we consider how the propagation of the photons in our method works in the case of multiple sources.

\subsection{Multiple sources and overlapping ionised regions}
\label{multi_source}
In the case of multiple sources in the same volume, the shells propagating from them will eventually overlap. In this section we discuss how we adapt the \ARTIST{} photon propagation in this case to obtain a self-consistent, photon-conserving ionisation fraction for these regions. Crucially, we aim to preserve photon directionality even in the case of overlapping ionised regions from different sources, which distinguishes \ARTIST\ from RT methods employing M1 closure~\citep[e.g.][]{Aubert2008}.

In order to conserve the directionality of  photon propagation in \ARTIST, photons in each cell need to be distinguished not only by the shell number $s$, but also by the source $l$ that originally emitted them. Photon packages inside cells will therefore be identified as $\gamma = \gamma_{l,s,d,i}$.  Leftover photons are added separately for each source, so that Eqn. \ref{gen_collect_d} becomes:
\begin{equation}
\gamma'_{l,s,d} = \sum_i \gamma'_{l,s,d,i}
\end{equation}
The redistribution of photons $\gamma'_{l,s,d}$ then takes place for each source \textit{separately} (again to ensure the correct directionality of photon propagation) using the method described in section \ref{asymmetric_prop}.

In the case of illumination by a single source, regions inside the cell illuminated by different shells are easily identifiable, as they are mutually exclusive. In the case of multiple sources illuminating the same cell, however, these become non-trivial to calculate as they are overlapping, and less relevant as photons are more likely to be spread across the cell. In order to save computational time, once shells within a particular cell start overlapping we no longer distinguish hydrogen atoms as belonging to different regions of the cell, and instead only calculate a unique ionisation fraction, assuming that all photons can potentially ionise all atoms within it. Eqn. \ref{leftNHI} therefore becomes:
\begin{equation}
N'_{\text{HI},i} = N_{\text{H},i} x_{\text{HI},i} - [\gamma_{i} - R_{\text{rec}}\text{d}t (1-x_{\text{HI},i})]
\end{equation}
where 
\begin{equation}
\gamma_{i} = \sum_l \sum_s \sum_d \gamma_{l,s,d,i}
\end{equation}
Notice that this value corresponds to the photoionising emissivity of the cell at that time-step.

The final neutral fraction in the cell will then be:
\begin{equation}
x'_i = \frac{N'_{\text{HI},i}}{N_{\text{H},i}}
\end{equation}
Since we don't compute the ionisation processes individually for each $\gamma_{l,s,d,i}$, we only obtain an overall number of leftover photons for the entire cell, which contains no information on which $l,s,d$ photon components had emitted them. Hence, $\gamma'_{i}$ is simply calculated as:
\begin{equation}  
\label{photons_left_multi}  
\gamma'_{i} =
\begin{cases}
|N'_{\text{HI},i}| & \text{ if } N'_{\text{HI},i} < 0 \\
0 & \text{ if } N'_{\text{HI},i} \geq 0 \\
\end{cases}
\end{equation}

In order to approximate the number of leftover photons that we expect a given photon component to produce, we assume that each $l,s,d$ component receives a fraction of the leftover photons which is equal to what was their original relative contribution to the ionising photons $\gamma_{i}$, i.e.
\begin{equation}
\label{leftovermulti}
\gamma'_{l,s,d,i} = \gamma'_{i} \times \bigg(\frac{\gamma_{l,s,d,i}}{\gamma_i} \bigg)
\end{equation}

Because of the number of separate components that need to be stored and computed in a cell illuminated by multiple sources, as the number of these increases the code becomes more and more computationally expensive, both in memory and time performance. To mitigate this in a practical but still physically accurate way, we introduce a ``background propagation" approach to evolve the diffuse photon background field, which we describe next.

\subsection{Background propagation}
\label{background}

The final module in \ARTIST\ handles the case where many photon wavefronts have overlapped, and the contribution from the sources within a given cell is sub-dominant compared to its external illumination.  In this case, we adopt an approximation for background photon propagation, with an adaptive criterion to decide when photons become part of the background.

The ionisation evolution of each cell is represented by the ($l,s,d,i$) components of photons that provide the ionising photon field $\gamma_i$. Storing and computing components that provide a negligible contribution to $\gamma_i$ is thus an unnecessary investment of computational memory.  We therefore introduce the free parameter $\epsilon$ to determine when a certain photon packet component ($l,s,d,i$) has a non-negligible contribution to the ionisation of a cell.  Non-negligible components in cell $i$ are  defined to be those that contribute -- in an ordered-sum over index $\mu$ from the largest to the smallest contributing component $\gamma'_{l,s,d,i}$ -- at least $\epsilon$ of the total leftover photons $\gamma'_{i}$. That is, once 
\begin{equation}
\frac{\sum_{\mu} \gamma'_{l,s,d,i}}{\gamma'_i} > \epsilon
\end{equation}
Components $l,s,d$ that do \textit{not} satisfy this criterion cease to be propagated through the previously described shell method, and instead are propagated via the background photon field. A value of $\epsilon = 0.01$ was chosen for the simulation based on convergence tests, which ensured that the background approximation yielded no visible effect on the output.

\begin{figure}
\centering 
\includegraphics[trim = 50 0 30 0, height = 6.0cm]{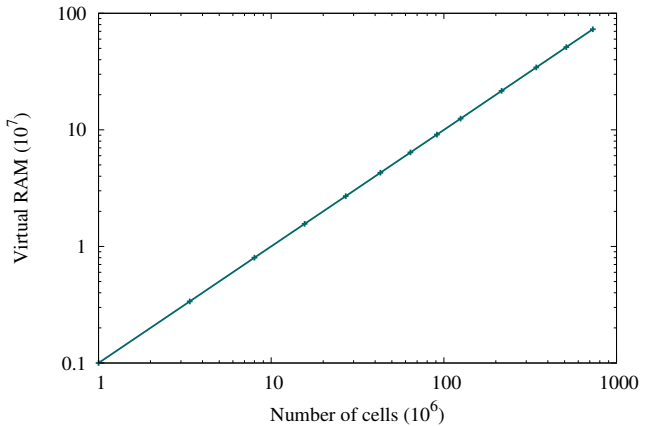}
\caption{{Virtual RAM required to include background photon propagation for cosmological grids with variable number of cells.}}
\label{bkgr_plot}
\end{figure}

The background field is evolved via a linear propagation of photons from any cell to its 26 nearest cell neighbours. 
When photons are added to the background, their directionality is chosen among the 26 available ones based on their original direction of travel from the source $l$ that emitted them (calculated from the position of cell $i$ relative to the host cell of source $l$). They are then propagated within the linear directions, losing all information on the source that emitted them and therefore shell $s$ and direction $d$ number, reducing the need to store this information for the next shell that they would have propagated into. In order to correct for the fact that the diagonal linear directions will propagate faster than the speed of light, we periodically ``freeze'' them (i.e. we do not propagate the photons stored in them at certain time-steps) to ensure that their propagation proceeds approximately at the speed of light. {The inclusion of these background linear channels for photon propagation of course comes at a RAM cost. In figure \ref{bkgr_plot} we illustrate this cost as a function of the number of cells considered in the simulation grid.}

\subsection{Summary of \ARTIST{} features}
\label{summary}
In this section, we summarise the main features of the algorithm. As shown in {section \ref{algorithm}}, \ARTIST: 
\begin{itemize}
\item Propagates radiation, on average, at the speed of light;
\item {Conserves  the  number  of  photons  in  the  simulation by construction;} 
\item Approximately conserves the directionality of photons as they propagate away from each source, {up to the point }when they are incorporated into the ionising 
background;
\item Allows for every cell to be partially ionised, i.e. $x_{\text{HI},i}$ can take any value between 0 and 1;
\item Computes  the  ionisation  state  of  each  cell  using  the cell's density, photo-emission, recombination rate and previous ionisation fraction {without any need for the averaging of these quantities over a sphere as required by ESF methods};
\item Self-consistently computes ionisation rates and photon propagation in cells illuminated by multiple sources;
\item Allows for variable degrees of angular resolution of the photon propagation around the source, and therefore reproduces shadowing and self-shielding effects with flexible accuracy;
\item Tracks  the  time-dependent  evolution  of  the  radiation field, and is hence applicable to on-the-fly simulations;
\item Introduces tunable approximations to ensure numerical tractability {can be optimised while ensuring the convergence of the physical result};
\end{itemize}

These features crucially distinguish \ARTIST\ from ESF methods which ``propagate" photons instantaneously, do not conserve the number of photons, do not account for photon directionality, only account for ionisations and recombinations averaged over an entire ionised region {, and make use of un-tunable approximations to optimise their computational requirements}. These improvements make \ARTIST\ a more physically accurate method for photon propagation.

Overall, \ARTIST{} provides a flexible and accurate evolution of the photon ionisation field for numerous RT applications.  As we discuss later, while these improvements incur additional computational cost relative to ESF methods, the requirements are still modest and feasibly allow evolving very large volumes at the required resolution for 21cm EoR studies. In the next section we discuss the performance of \ARTIST{} in standard RT tests.

\section{RT Tests}
\label{compRT}
In order to assess how \ARTIST{}  compares in accuracy to other RT methods used in cosmological reionisation simulations, we take advantage of the  ``Cosmological radiative transfer codes comparison project'' compiled by \cite{Iliev2006} to test its performance, and further perform direct comparisons versus the \cite{Finlator2018} cosmological radiative hydrodynamic simulations. In particular, we discuss \ARTIST's ability to:
\begin{itemize}
\item simulate a pure-hydrogen, isothermal HII region expansion (section \ref{stromgr});
\item account for self-shielding and shadowing effects in the case of over-dense regions in the HI medium (section \ref{shadowing}); 
\item reproduce the cosmic ionisation history of the 12 Mpc/h Technicolor Dawn cosmological rad-hydro simulations of  \citealt{Finlator2018} (section \ref{kristian});
\end{itemize}
The only test from \cite{Iliev2006} that we do not consider (Test 2) concerns the temperature-evolution of the IGM following photoionisation, as the temperature of the IGM is not tracked by our RT code.

\subsection{Pure-hydrogen, isothermal HII region expansion}
\label{stromgr}
We begin by considering the benchmark case of the propagation of photons in an isothermal medium of constant density, using the simulation parameters considered by \cite{Iliev2006} in their Test 1. We thus adopt a 13.2 kpc box, an ionising source at the centre with a photon-emission rate of $\dot N_{\gamma} = 5 \times 10^{48}$ photons s$^{-1}$, and assume a constant hydrogen number density of $n_{\text{H}} = 10^{-3}$ cm$^{-3}$. Given an assumed temperature of $T = 10^4 \text{K}$, the recombination rate is $\alpha_{\text{B}} = 2.59 \times 10^{-13} \text{cm}^3 \text{s}^{-1}$, with a recombination time of $t_{\text{rec}} = 3.86 \times 10^{15}$ s $= 122.4$ Myr.

Given these parameters, and assuming a thin transition region between the ionised and neutral part of the region around the source, the time evolution of the ionised front can be approximated by the following analytical solution:
\begin{equation}
\label{rI}
r_I = r_S[1-\text{exp}(-t/t_{\text{rec}})]^{1/3}
\end{equation}
where 
\begin{equation}
r_S = \bigg[\frac{3 \dot N_{\gamma}}{4 \uppi \alpha_B(T)n_H^2} \bigg]^{1/3}
\end{equation}
In the comparison study of the different RT codes, the ionised radius $r_I$ is chosen to be the one at which 50\% of the material has been ionised. As discussed in \cite{Iliev2006}, this choice is rather arbitrary and can lead to small differences between different RT methods.

\begin{figure}
\centering 
\includegraphics[trim =6 59 0 95, width = 8cm]{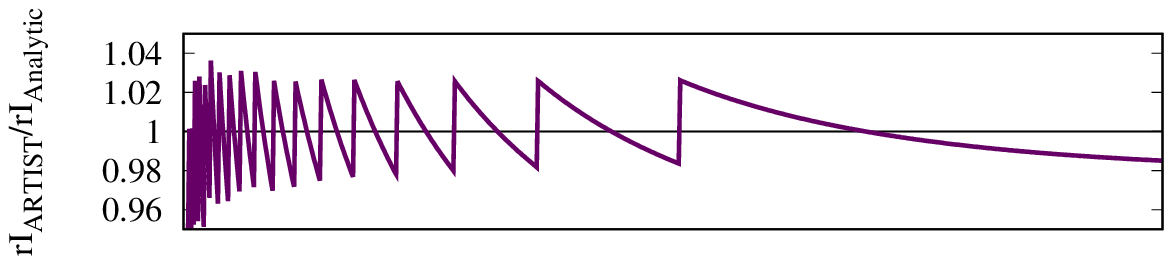} \\
\includegraphics[trim = 0 0 0 50, width =8cm]{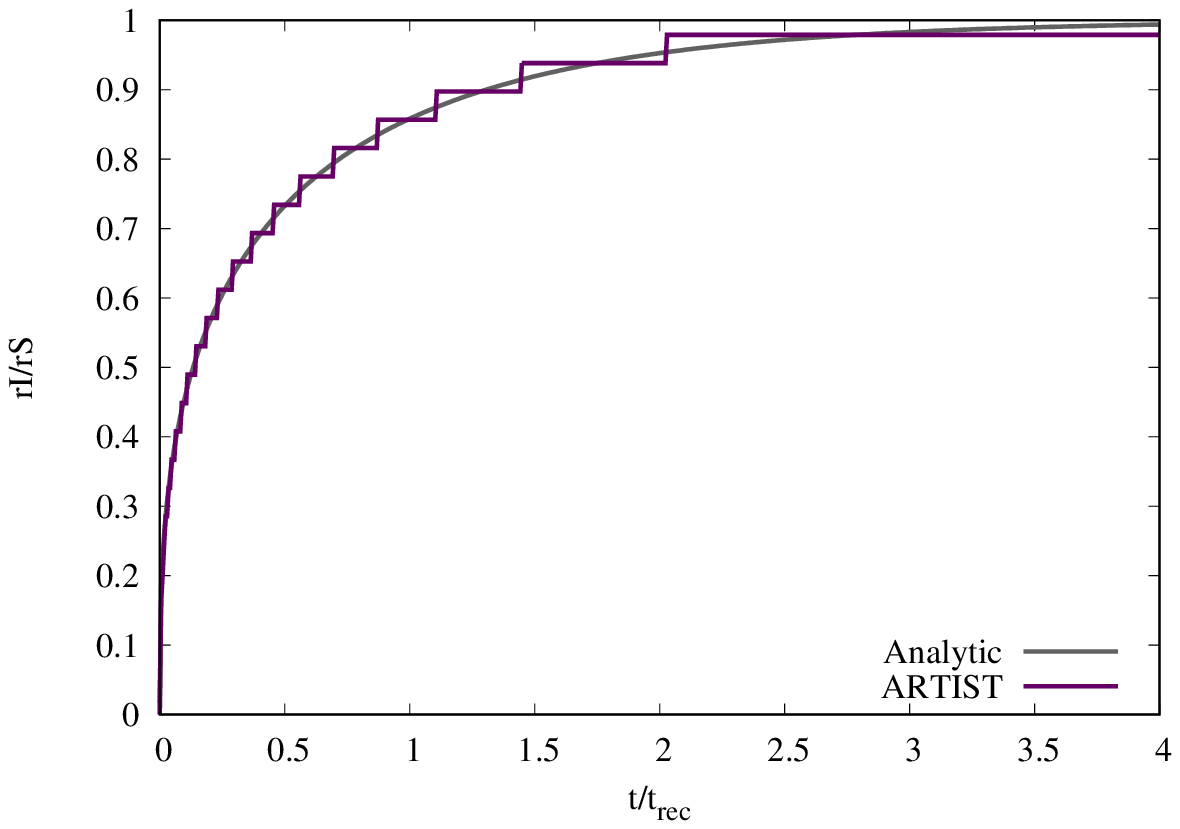} \\
\caption{Comparison between the radius of the ionised region $r_\text{I}$ in units of the Str\"omgren radius $r_\text{S}$, as a function of time, calculated using \ARTIST{} and the analytic solution described Eqn. \ref{rI}. The ratio between the two is shown in the top plot. As shown by this figure, \ARTIST{} closely follows the analytic result. The range of its ratio is also within the one spanned by other cosmological RT methods (see Fig. 7 in \citet{Iliev2006} for a direct comparison).}
\label{rI_rS_fig}
\end{figure}

Fig. \ref{rI_rS_fig} compares the evolution of the radius of the ionised region with the equivalent analytic result until the source reaches the Str\"omgren radius at $t/t_{\text{rec}} \sim 4$. The jagged nature of the ionisation front is a consequence of our Cartesian grid. Each step oscillates about the analytic solution, with an amplitude dependent on the chosen grid resolution. The numerical error incurred by our method, shown by the top plot in the same figure, is well within the range of error (see Fig. 7 in \cite{Iliev2006} for a direct comparison) incurred by other methods by the time the ionised region has reached its full spatial extent.  

\begin{figure}
\centering 
\includegraphics[trim = 50 0 30 0, height = 7.0cm]{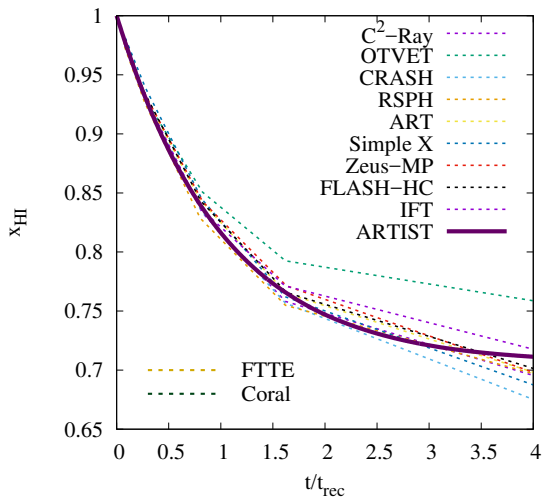}
\caption{Evolution of the average neutral fraction in the simulation volume as calculated by \ARTIST{} \textit{(thick bordeaux line)}, compared with those found by the cosmological RT codes considered in \citet{Iliev2006} -- where references to these codes can be found. Our results are highly consistent with those of other RT methods, and follows the ionisation fraction evolution found by the majority of methods.}
\label{average_ion_test}
\end{figure}

Fig. \ref{average_ion_test} shows the evolution of the average neutral fraction over the same time range. Notice that, compared to the previous test which only considered the distance of the ionising wavefront from the source, the average neutral fraction contains additional information on the ionisation structure within the ionised region. As seen in Fig. \ref{average_ion_test}, our method is in agreement with the large majority of the methods considered.  \ARTIST{} therefore straightforwardly passes this simple Str\"{o}mgren sphere test.

\subsection{Self-shielding and shadowing effects}
\label{shadowing}

   \begin{figure}
   \label{clump_shielding}
   \centering
   \includegraphics[trim=20 0 0 0, height=7cm]{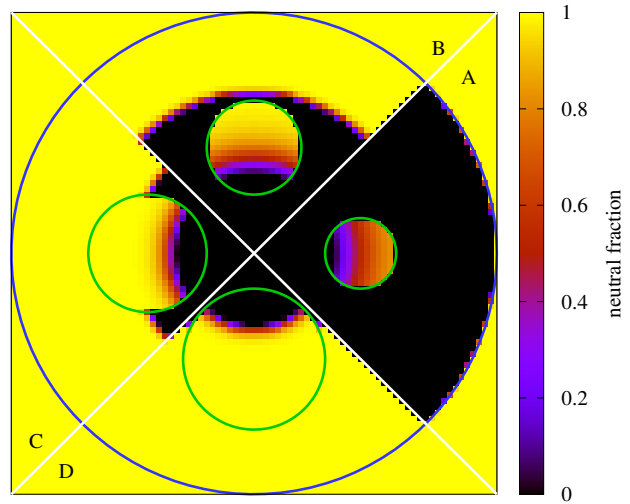} 

      \caption{Evolution of the ionisation front in the presence of an intervening over-dense clump. The four different quadrants - indicated by the white lines - illustrate how this is computed by \ARTIST{} in the case  of a clump (shown by the green circles) of fixed physical but resolved with increasingly high angular resolution (A to D). The blue circle indicates the position of the ionisation front in the worst-case resolution of the clump. The constant physical size of the clumps is ensured through the rescaling of the luminosity emitted by the source in each quadrant, so that the flux reaching the clump is always constant.  }
         \label{shadowing_solid_angle}
   \end{figure}

In this section we consider the case of an anisotropic density distribution, and discuss \ARTIST{}'s ability to reproduce self-shielding and shadowing effects in comparison both to ESF and other RT methods.

ESF techniques estimate the evolution of the ionisation fraction by assuming values for the recombination rate and density which are averaged over all cells included within the spherical volume selected (e.g. see \citealt{Santos2010,Hassan2017}). Crucially, the propagation of photons from the point of emission to the cells in which the ionisation process takes place is not computed in a self-consistent way, by the very nature of this algorithm. \ARTIST, on the other hand, tracks the evolution of the ionisation state of hydrogen in the simulation for each cell independently, using the cell's individual recombination rate, density, and ionisation state (which, unlike in ESF-based simulations, can be partial in cells other than the one containing the source). The photon propagation up to those cells is also accounted for self-consistently.

The photon propagation by \ARTIST{} is affected in the following way by the presence of density inhomogeneities. As discussed in section \ref{asymmetric_prop}, the leftover photons from all cells illuminated by the shell-section ($s,d$) are added together (see Eqn. \ref{gen_collect_d}) before being redistributed among cells illuminated by the next shell-section ($s+1,d$), as shown in Eqn. \ref{gen_redistr_d}. In the case of a homogeneous density distribution, this will be perfectly accurate. If, however, the density of the cells is
illuminated by ($s,d$) is inhomogeneous, the cells in shell-section ($s+1,d$) will receive an amount of photons obtained from the averaging of the (fewer) leftover photons from higher density cells, and the (more numerous) ones from lower-density cells (see section \ref{asymmetric_prop}). 

This leads to a smearing of the asymmetric effect that over-dense regions should have on the propagation of the ionisation front. The level of inaccuracy introduced therefore depends on the angular resolution of these inhomogeneities, or in other words the fraction of the shell-section ($s,d$) that these inhomogeneities occupy: if the over-dense cells occupy a small fraction of the shell-section ($s,d$), the cells in ($s+1,d$) located beyond them (which in principle should receive very few photons) will be `contaminated' by the many left-over photons from the lower density cells in ($s,d$). Besides the angular resolution of inhomogeneities, the accuracy of the photon propagation is also affected by the density distribution in the next shell, and in particular the density of cells in ($s+1,d$) located beyond the higher-density cells in ($s,d$), in the following way:
\begin{itemize}
\item If the cells in ($s+1,d$) beyond the higher-density cells in ($s,d$) are themselves high-density, they can partially correct for the `contamination' from lower density cells in ($s,d$) if they are dense enough to absorb the excess photons without a significant effect on their ionisation fraction. This implies that the self-shielding effect can still be accounted for, albeit with a variable level of inaccuracy. 
\item If the cells in ($s+1,d$) beyond the higher-density cells in ($s,d$) are of low density, their ionisation fraction will be strongly affected by the excess photons resulting from the averaging process, resulting in the smearing of the shadowing effect. However notice that the presence of a clump in direction $d$ will still result in a slower propagation of the overall wavefront in that direction compared to those without high-density regions, as fewer photons overall will be propagated to shell ($s+1,d$). The algorithm therefore will still roughly account for one of the shadowing-effect features.
\end{itemize}
The accuracy of \ARTIST{} in reproducing these two effects is therefore dependent on the angular resolution of the overdensity, as shown in Fig. \ref{shadowing_solid_angle}. Again referring to \cite{Iliev2006}, we consider a simulation box with a source at its centre and hydrogen density $n_{\text{H}} = 2 \times 10^{-4}$. We propagate the photons emitted by the source in four directions, and include an over-dense clump ($n_{\text{clump}} = 0.04 \text{ cm}^{-3} = 200 ~n_{\text{H}}$) in each of them. In order to ensure that the flux reaching the clump is the same, but that the angular resolution considered by \ARTIST{} is different, we vary the clump's radius  ($r_{\text{clump}}= 0.792, 1.056, 1.32$ and $1.58$ kpc), while keeping it at a constant distance from the source of $2.38$ kpc, and rescale the luminosity reaching the clumps by their angular size. Note that this test is exactly equivalent to increasing or decreasing the angular resolution around a clump of constant size and position.

Fig. \ref{shadowing_solid_angle} shows the propagation of the ionisation front in the presence of a clump illuminated by a constant flux with four different angular resolutions. The snapshot is taken at the time when the least resolved direction reaches the edge of the simulation box. 

With this plot we illustrate the following: as the covering factor of the clump increases (A to D), the ability of the algorithm to account for the self-shielding effect improves, as shown by the fact that the average ionisation fraction inside the clump decreases. Even with the lowest possible resolution (quadrant A), the self-shielding is partially accounted for by the use by \ARTIST{} of local densities to estimate the ionisation fraction. As the angular resolution decreases (D to A) the shadowing is `smeared' by the propagation of photons behind the clump. The shadowing effect is however still somewhat accounted for by the slower ionisation of regions behind higher-density ones.

From the figure, one can see how in the case where the clump occupies most of the solid angle, virtually no photons are propagated further, correctly accounting for both self-shielding and shadowing. Any higher angular resolution of over-dense regions, e.g. in the case where a single clump is split across several pyramids, is therefore unnecessary from a shadowing-accuracy point of view. As the clump's resolution is reduced, the number of photons escaping the lower density regions increases and these start to contaminate the cells located beyond the overdensity. Notice however that:
\begin{itemize}
\item Thanks to the use of the local density to calculate the ionisation fraction, self-shielding effects are still visible regardless of the clump resolution;
\item The propagation of the wavefront in the directions where the clump is less resolved is still significantly delayed, since fewer photons are transmitted through; so although the morphology of the shadowing may be blurred, regions behind over-densities will still be ionised more slowly than those not shielded by clumps.
\end{itemize}

In summary, the accuracy of \ARTIST{} in reproducing the self-shielding and shadowing effects improves as the over-dense region occupies higher fractions of the shell cross-section, or equivalently as the angular size of the clumps increases. This is self-evident when thinking of \ARTIST{} as a solid-angle averaging of a Monte-Carlo-propagated photon package. 

Thanks to its use of local densities, recombination rates and ionisation states to account for photon absorption, \ARTIST{} can nonetheless partially compensate for these inaccuracies in lower angular resolutions, and therefore still roughly reproduce both self-shielding and shadowing effects.

\subsection{Direct comparison with a multi-source, cosmological RT simulation}
\label{kristian}

   \begin{figure*}
   \centering 
        \includegraphics[trim = 60 0 73 0,clip,height=5.6cm, width=5cm]{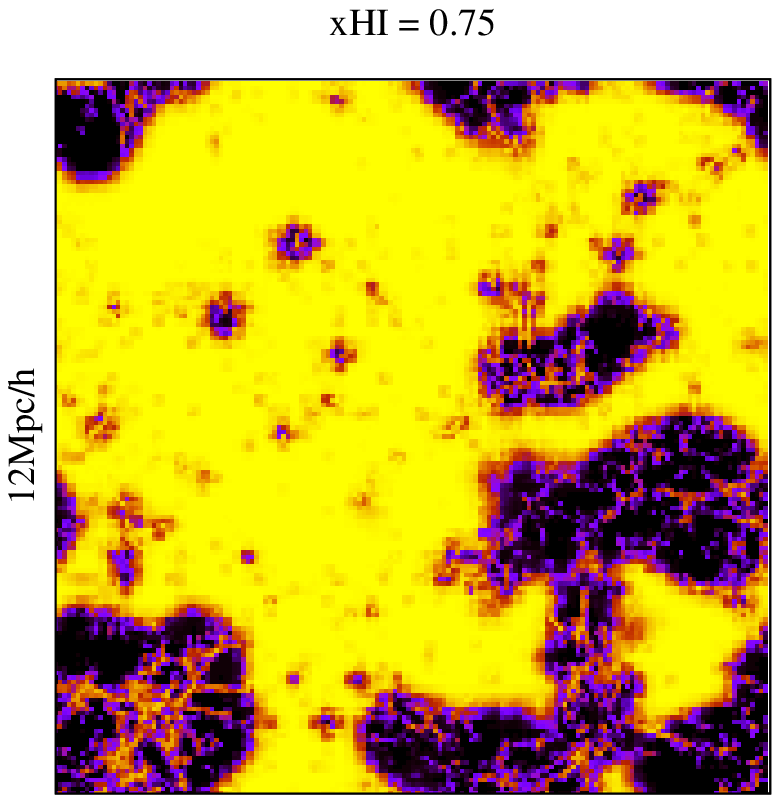}
          \includegraphics[trim = 68 0 80 0,clip,height=5.6cm]{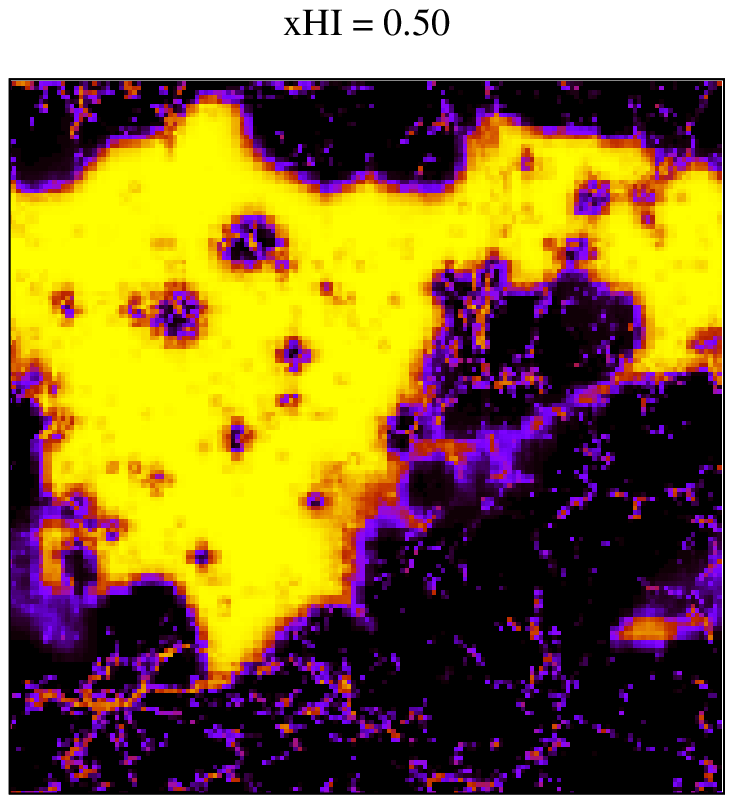}
             \includegraphics[trim = 50 0 50 0,clip,height=5.6cm]{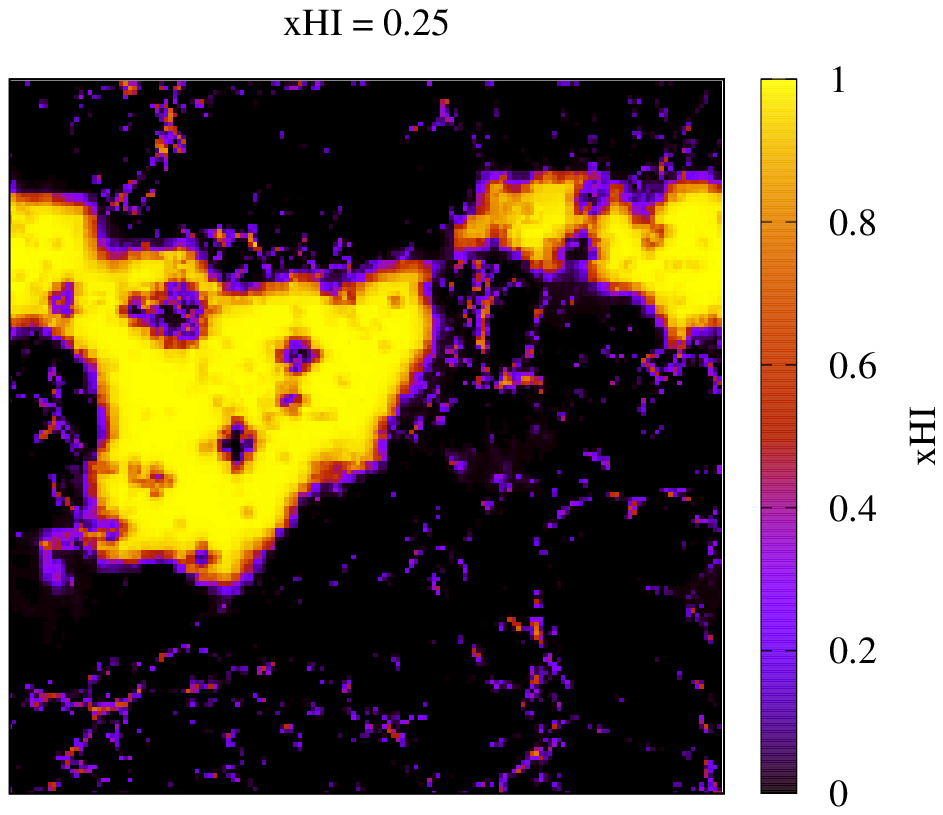}  
       \includegraphics[trim = 60 0 78 33,clip,height=5cm, width=5cm]{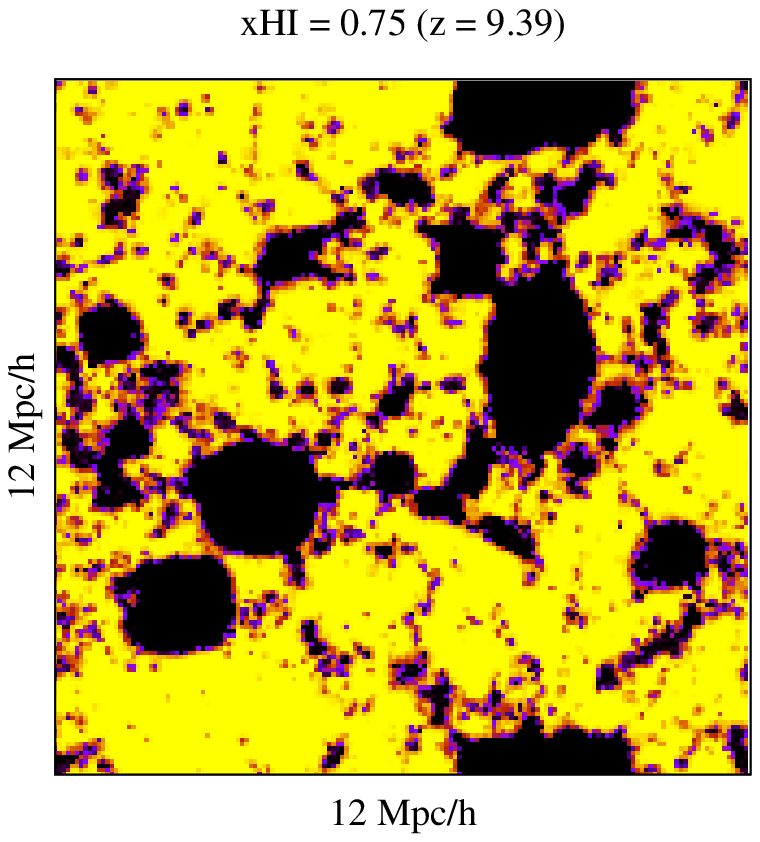}
          \includegraphics[trim = 69 0 84 33,clip,height=5cm]{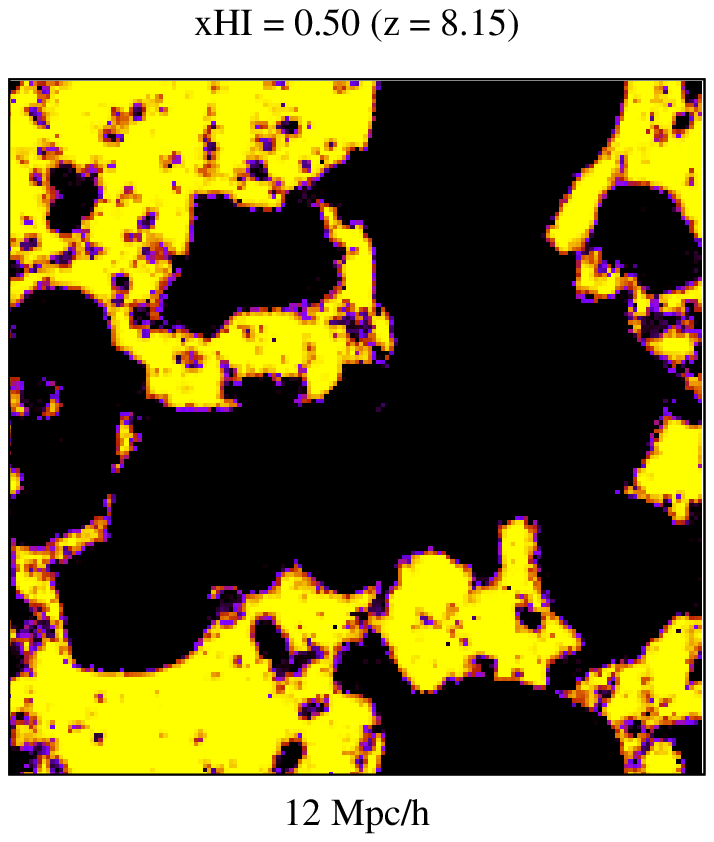}
             \includegraphics[trim = 50 0 54 33,clip,height=5cm]{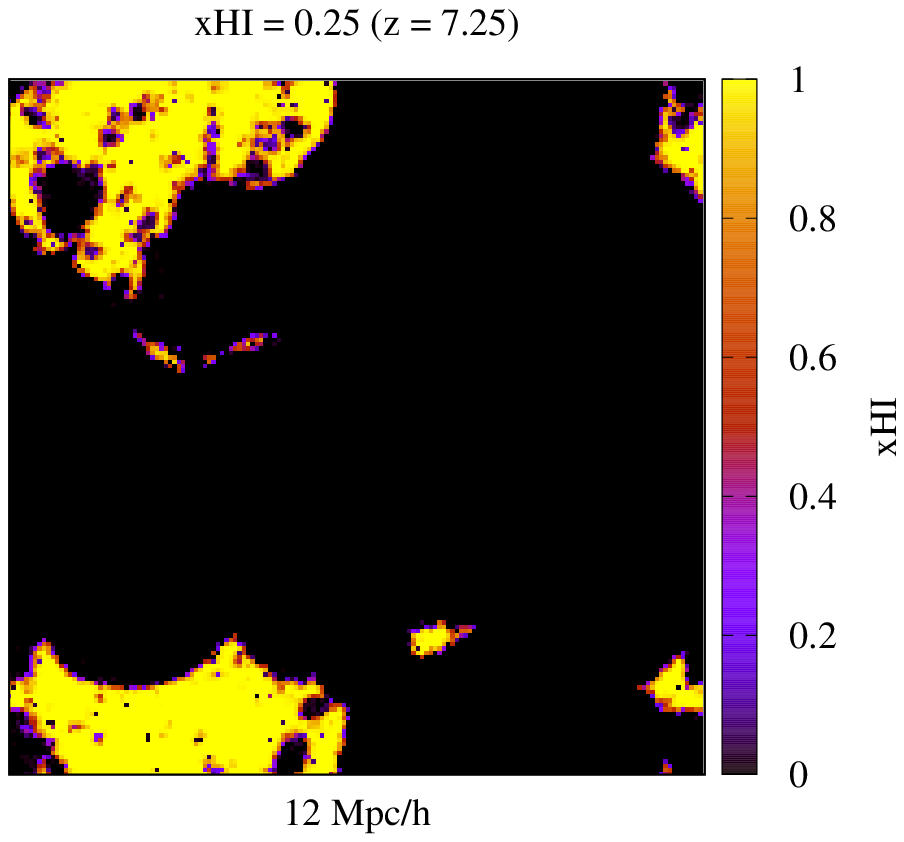}
          
          \caption{{Ionisation maps for the 12 Mpc/h simulation considered in comparison with TD at an average neutral hydrogen value of $x_{\text{HI}} = $0.75 \textit{(left)}, 0.50 \textit{(centre)} and 0.25 \textit{(right)}.} {The top row shows the maps from the TD simulation, and the bottom one the ones from \ARTIST{}.}}
                \label{fig:Fin_maps}
   \end{figure*}

   \begin{figure}
   \centering
      \includegraphics[trim = 50 0 50 0, height=6cm]{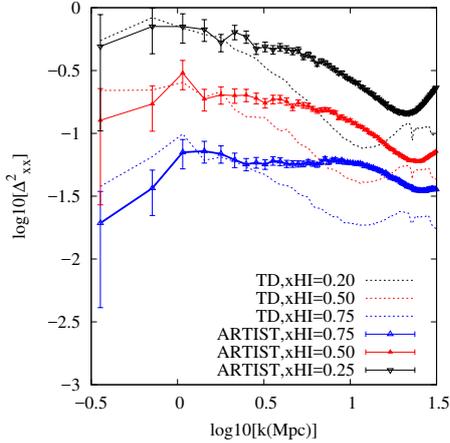}\\      
    
            \caption{{Ionised hydrogen power spectrum $\Delta^2_{\text{xx}}$ shown at three values of the neutral fraction ($x_{\text{HI}}$ = 0.75, 0.50, and 0.25) for the 12Mpc/h simulation box considered in section \ref{kristian}. {The error bars show the uncertainty due to cosmic variance, as defined in the text.}}}
               \label{fig:Fin_PP}
   \end{figure}  
   
   \begin{figure}
   \centering
  \includegraphics[trim = 10 70 0 50,height=4.2cm, width = 9cm]{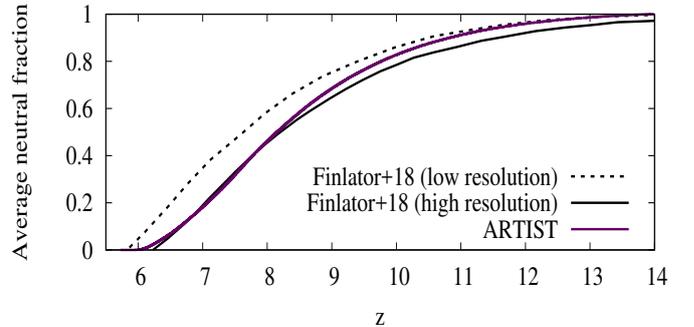}
      \caption{Evolution of the average neutral fraction in a 12Mpc/h box as obtained by \ARTIST{} {(purple line)}, obtained assuming the escape fraction function described in \ref{fesc_gal} with fitting parameters $A = 1.95$ and $f_{\text{esc,max}} = 0.36$. 
      \ARTIST{} much more closely reproduced the high-resolution evolution found by the moment-based RT in \citet{Finlator2018} (shown in the black solid line).}
         \label{Average_ion_fin_comp}
   \end{figure}

In this section we carry out a direct, side-by-side comparison with a state of the art cosmological radiative hydrodynamical code, namely \citep{Finlator2009}'s custom RT version of GADGET-3, which evolves the galaxy population as sources and the RT on an overlaid grid using a moment-based method with the Eddington tensor periodically computed via long characteristics. In particular, we seek to reproduce the recent results from the Technicolor Dawn (TD) Simulation \citep{Finlator2018}, a 12 Mpc/h cosmological simulation including galaxy formation physics and multi-frequency RT evolved concurrently, by replacing its photon propagation algorithm with \ARTIST{}. {In doing so we assess the accuracy of \ARTIST{} in the presence of multiple sources, and hence in the case of overlapping ionised regions. The comparison with the moment-based RT method used in the TD simulation can be generalised to other RT methods by making use of the method's original reference \citep{Finlator2009}}

In order to do so, we require ionising photon emission rates, recombination rates and density distributions. We obtain these quantities for our \ARTIST{} run indirectly from the Technicolor Dawn simulation itself. Specifically, following the procedure described in \cite{Hassan2016}, we used parameterised relations for the ionisation and recombination rates as a function of halo mass and local density, respectively. For the recombination rate, we update the parameters of the fitting function to ensure these are accurate for the spatial resolution of our simulation (0.075 Mpc/h). Given the same fitting function as a function of overdensity $\Delta$ and redshift $z$:
\begin{equation}
\frac{R_{\text{rec}}}{V} = A_{\text{rec}} \times (1 + z)^{D_\text{rec}} \bigg(\frac{(\Delta/B)^{C_{\text{rec}}}}{1+(\Delta/B)^{C_{\text{rec}}}} \bigg)^4
\end{equation}  
we update the fitting parameters to take the following values: $A_{\text{rec}} =22.51 \times 10^{-24}$ cm$^{-3}$ cm$^{-1}$, $B_{\text{rec}} = 2.69$, $C_{\text{rec}}=0.81$, $C_{\text{rec}}= 5.13$. 

We use identical initial conditions to Technicolor Dawn, re-gridded to the resolution for \ARTIST{} (N = 160) and evolve the
density field using {\sc SimFast21}.  For a more detailed discussion of how \ARTIST{} is incorporated in {\sc SimFast21}, see \ref{recrioN_subsec}. We further assume the same cosmological parameters and escape fraction used in \cite{Finlator2018}. The latter is the redshift-dependent escape fraction evolution described in Eqn. 7 of their paper, parametrised as:
\begin{equation}
\label{fesc_gal}
f_{\text{esc,gal}}(z) = 0.176 \text{ }\bigg(\frac{1+z}{6} \bigg)^{A }
\end{equation}
with a cut-off escape fraction of $f_{\text{esc,max}}$. 

The TD Simulation considers two different resolutions for their radiative transfer solver, a low-resolution ($32^3$) and a high-resolution one ($64^3$). Due to the nature of their RT method, the number of photons per hydrogen atom emitted in the simulation must be adjusted in order to compare the two consistently. In particular, this leads to a need for an artificial increase in the escape fraction when accounting for more highly refined - and therefore more accurate - grids. As our RT doesn't require this rescaling, an artificial increase in the number of photons would make a comparison with the more accurate higher-resolution TD Simulation inconsistent. We therefore adopt the original $f_{\text{esc,gal}}$ {parameters chosen prior the increase in the resolution ($f_{\text{esc,max}}=0.36$, $A=1.95$), to compare} \ARTIST{} to the higher-resolution Technicolor simulation.

{Fig. \ref{fig:Fin_maps} shows the ionisation maps of the simulation box at an average neutral hydrogen fraction of $x_{\text{HI}}$ = 0.75, 0.50, and 0.25, whereas Fig. \ref{fig:Fin_PP}} shows the corresponding power spectra in the form defined in Eqns \ref{eqn:Dxx}. {The plot includes the error bars due to cosmic variance, calculated as $\sim P(k)/\sqrt{N_{\text{err}}}$ where
}

\begin{equation}
     N_{\text{err}} = 4 \uppi k^2 \text{d}k_{\text{bin}}/\text{d}k^3,
     \end{equation}
{where $\text{d}k_{\text{bin}}$ is the bin size, and $dk=2\uppi/L$ with $L$ being the size of the simulation box. The inclusion of the cosmic variance effect clearly explains the discrepancy at larger scales of the power spectra.  The discrepancy at smaller scales, on the other hand, can be explained by the fact that \ARTIST{} is first-order accurate in space and has no sub-grid complexity, and uses a higher grid resolution than TD. Furthermore, as it is a monochromatic and non-diffusive RT solver, the I-fronts are rather sharper and smaller-scale features are resolved, boosting small-scale power in \ARTIST{}. The remaining discrepancy therefore reflects an advantage in accuracy that \ARTIST{} has over the second-order and diffusive RT method used in TD.}

In Fig. \ref{Average_ion_fin_comp}, we verify that \ARTIST{} reproduces very closely the evolution of the average neutral fraction found by the highest-resolution simulation (solid black line) of the Technicolor Dawn Simulation down to $z \sim 6$.  This confirms that \ARTIST{} can reproduce quite well -- given the adoption of approximate parametrised relations, and the less sophisticated treatment of frequency-dependent photon propagation -- the evolution of the global ionisation fraction in the case of multiple sources found by moment-based RT methods. 

Technicolor Dawn used $\sim$40,000 CPU hours for their high-resolution RT simulation, approximately half of which went to the RT.  \ARTIST{}, by contrast, employed less than 2,000 CPU hours and was run on a single workstation, making it at around 10 times faster than TD (neglecting factors such as processor speed and network topology).  Note that the improvement in computational performance of \ARTIST{} owes to both the use of a semi-numerical approximation for ionising photon emission and recombination rates \textit{and} our faster radiative transfer approach. Importantly, {although only considering a single frequency,} \ARTIST{} evolved a significantly higher resolution grid than TD's $64^3$; it is currently infeasible to evolve a $160^3$ RT grid using the full radiative hydro approach in TD. This means that \ARTIST\ can treat significantly higher dynamic range for modest computational cost. {An additional consideration to make is that, due to the fact that the redshift step in \ARTIST{} is constrained by the cell size, when considering such a small volume an extremely large number of redshift steps will be selected in the $z = 6-14$ range, namely $\sim 12,700$. This disadvantage however decreases as the simulation volume increases, which therefore optimises \ARTIST{} for the large-scale simulations of the EoR. This is discussed further in section \ref{performance}.}
\newline

In this section we have shown via benchmark tests and a direct comparison with the results of the Technicolor Dawn rad-hydro simulation that \ARTIST{} is competitive in accuracy when compared to other cosmological RT methods. Thanks to the fact that one of its defining characteristics is its optimised computational time efficiency (see  section \ref{performance}) \ARTIST{} significantly enhances the accuracy of semi-numerical codes, currently the only method capable of predicting EoR 21cm power spectra on very large scales, in approximating full radiative hydrodynamic simulations when simulating the ionisation history of the EoR.  \newline

In the next section, we quantify the difference that replacing the ESF with \ARTIST{} in such codes makes when making predictions of the large-scale EoR signal as will be observed by future radio experiments.

\section{EoR Simulation With \ARTIST}
\label{esf}

In order to quantify the improvement introduced by our RT method over ESF in the study of the large-scale EoR signal, we discuss the difference in the output of the latest version of our in-house semi-numerical code {\sc SimFast21} \citep{Hassan2017} when replacing ESF with \ARTIST. \ARTIST's implementation will be compared to {\sc SimFast21}'s native, instantaneous ESF (from now on InstESF) {originally discussed in} \cite{Mesinger2007}, {which makes use of a whole-sphere flagging method}.

Initial conditions are set up in a cosmological box of size $L = 75$ Mpc with N$ = 160$ cells per side, which allows us to obtain a spatial resolution of 0.468 Mpc. We evolve the simulation from $z$ = 14 to $z$ = 5. We assume the same escape fraction selected by \cite{Hassan2017}, $f_{\text{esc}} = 0.25$. The assumed cosmology is a $\Lambda_{\text{CDM}}$ cosmology with $\Omega_{\text{M}} = 0.3$, $\Omega_{\Lambda} = 0.7$, $h \equiv \text{H}_0/(100 \text{km/s/Mpc}) = 0.7$, a primordial power spectrum index $n_s = 0.96$, an amplitude of the mass fluctuations scaled to $\sigma_8$ = 0.8, and $\Omega_{\text{b}} = 0.045$.
 
In the next section, we discuss how \ARTIST{} is implemented in {\sc SimFast21}. In sections \ref{av_fraction} to \ref{power_spectrum} we compare the evolution of the ionisation history during the EoR as obtained by InstESF and \ARTIST. In particular, we discuss the evolution of the average neutral fraction (section \ref{av_fraction}), its morphology (section \ref{morphology}), and power spectrum (section \ref{power_spectrum}). Finally in section \ref{summary_comp_sifast21} we summarise the main findings of this comparison study.

Since our aim is to quantify the difference in the results obtained using these two methods, we refrain from speculating on the physical validity of simulation parameters based on a direct comparison with observations.

\subsection{Implementing \ARTIST{} in {\sc SimFast21}}
\label{recrioN_subsec}
As previously discussed, \ARTIST{} requires the following input parameters in each time-step for each cell in the grid: the ionising photons emission rates ($R_{\text{ion}}$), the total recombination rates ($R_{\text{rec}}$), and the number  of hydrogen atoms ($N_{\text{HI}}$). 
Using {\sc SimFast21}, we obtain snapshots of these quantities at redshift intervals d$z_{\text{snap}} = 0.025$, and assume these to be constant over that redshift interval. This approximation is verified to be negligible by performing convergence tests over the variable d$z_{\text{snap}}$. Notice that d$z_{\text{snap}}\neq \text{d}z$, since the latter is constrained by the cell size $\text{d}x$ to be: 
\begin{equation}
\text{d}z = (\text{d}x/c) \times H(z)\times(1+z),
\label{dzstep}
\end{equation}
where $H(z)$ is the Hubble function, as discussed in Eqn.\ref{timestep}.

 Although this particular  application considers the post-processing of the density field, \ARTIST{} remains an approach which is fully implementable in self-consistent simulations of the matter and radiative fields, due to the fact that the input parameters are updated at every time-step. 

The semi-numerical simulation {\sc SimFast21} applies a Monte Carlo Gaussian approach to generate the dark matter density field in the linear regime, and then dynamically evolves it into a non-linear field using the \cite{Zeldovich1970} approximation. Dark matter (DM) halos are then identified using an excursion set formalism, which collapses a given region into a halo if its mean overdensity is higher than a given threshold (see \citealt{Mesinger2007} and \citealt{Santos2010} for more details) and setting a minimum halo mass of $10^8 \text{M}_{\odot}$.
From this density distribution, together with the redshift evolution of the cell size in the cosmology assumed, we can therefore obtain the number of hydrogen atoms $N_{\text{H}}$ in each cell.

From this density and DM halos distribution, semi-numerical simulations assume a relation for the recombination rate and photoionising emission. As discussed in section \ref{kristian}, the version of {\sc SimFast21} considered in this comparison was the first such simulation to rely on high-resolution, hydrodynamical simulations of smaller cosmological volumes \citep{Finlator2015}, and combined with a larger hydrodynamical simulations \citep{Dave2013}, to obtain more physically-grounded parametrised relations as a function of the DM halo mass \citep{Hassan2016}. 

The parametrised relations for $R_{\text{ion}}$ assumed are as follows:
\begin{equation}
R_{\text{ion}} = M_\text{h} \times A_{\text{ion}} (1+z)^{D_{\text{ion}}} (M_\text{h}/B_{\text{ion}})^{C_{\text{ion}}} \text{exp}[-(B_{\text{ion}}/M_\text{h})^3]
\end{equation}
where $A_{\text{ion}} = 1.08 \times 10^{40} \text{M}_{\odot}\text{s}^{-1}$, $B_{\text{ion}} = 9.51 \times 10^7 \text{M}_{\odot}$, $C_{\text{ion}} = 0.41$ and $D_{\text{ion}} = 2.28$. 

Recombination rates assumed in this section, unlike those of section \ref{kristian}, are simply obtained from the hydrogen density squared of each cell, and assuming a case B recombination rate of $\alpha_B = 2.6 \times 10^{-13} \text{cm}^3 \text{s}^{-1}$, as relevant for hydrogen at gas temperature $T = 10^4$ K. Although more sophisticated methods for recombination rate estimation are available (see \citealt{Raicevic2011, Sobacchi2014,Hassan2017}) the choice of recombination method has no impact on the comparison between InstESF and \ARTIST, given that it is the same for both cases. In this section we therefore opt for the simplest approximation.

\subsection{Average neutral fraction evolution}
\label{av_fraction}
\begin{figure}
   \centering
   \includegraphics[trim = 50 0 50 0,height=6.5cm]{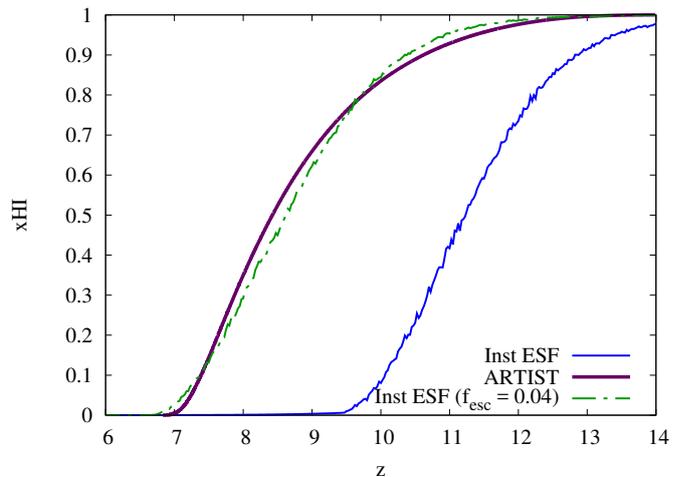}   
      \caption{Evolution of the average neutral fraction $x_{\text{HI}}$ in the 75 Mpc simulation box.  \ARTIST{} is shown to produce a significantly slower ionisation history ($\tau$ = 0.067) than the InstESF ($\tau$ =0.097). This results in an overall later conclusion to the ionisation process than by the ESF methods. In order for InstESF to match \ARTIST{} the escape fraction would have to be rescaled down to $f_{\text{esc}}=0.04$.
              }
              
         \label{average_ionisation}
   \end{figure}

In this section we discuss the evolution of the average neutral hydrogen fraction in the simulation box, as obtained by the two ESFs and \ARTIST.

Figure \ref{average_ionisation} shows that \ARTIST{} produces a significantly different evolution of the average neutral fraction compared to InstESF. 

The ionisation of the IGM is completed far earlier by InstESF ($\tau$ =0.097). Indeed, the escape fraction in the InstESF case has to be reduced to $f_{\text{esc}} = 0.04$ (from $f_{\text{esc}} = 0.25$) in order to obtain a redshift evolution similar to \ARTIST{}; this is the escape fraction found in \citet{Hassan2016} that was required to match EoR data using the InstESF code. The need for $f_{\text{esc}}$ {to be rescaled to lower values in order to match whole-sphere flagging ESF methods to RT simulations confirms the finding of other comparison studies} (see \citealt{Hutter2018} and references therein for a recent review). {The same study also highlights how their single-cell flagging ESF method removes the need for such $f_{\text{esc}}$ rescaling, although not being able to solve the effect of lack of photon conservation (see discussion in section \ref{power_spectrum}).}

The discrepancies between the two methods can be explained as follows. The slower reionisation of the IGM found by \ARTIST{} compared to the ESF method is most likely due to the issues faced by the latter to conserve the number of photons in the simulation - particularly in overlapping ionised regions - resulting in a possible overestimation of the number of photons responsible for ionising the IGM. A more physical, number-conserving propagation of the photons should therefore slow the ionisation process, as found by \ARTIST. 

The importance of partial ionisation may be partly an artefact of low resolution, because EoR ionisation fronts are in fact expected to be only around 10kpc in width \citep{daloisio2019}, and hence a simulation with sufficiently high resolution should have few partially ionised cells. However, at the resolutions that are required for large-scale EoR runs (including the tests presented here), the ionisation fronts' widths will be exaggerated, leading to a larger volume-fraction of the IGM that is partially ionised. {We notice that this appears to be in contradiction with the findings of \cite{Zahn2011}}. The impact of this limitation on small-scale features is difficult to assess directly, but it should be weak on large scales. For this comparison, we therefore focus on comparing the two methods at a constant spatial resolution. 

In the next section, we look into more detail at the morphology of the ionised regions for the different cases.

\subsection{Ionisation morphology}
\label{morphology}

\begin{figure*}

   \centering

     \includegraphics[trim = 135 0 25 0, height=6.3cm]{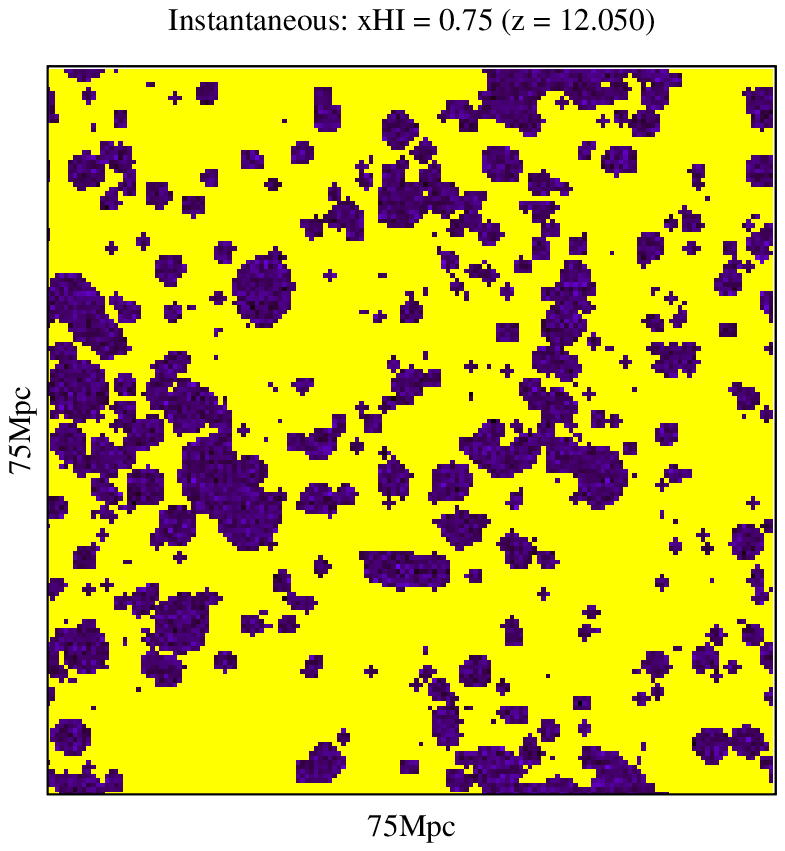} 
   \includegraphics[trim = 100 0 50 0, height=6.3cm]{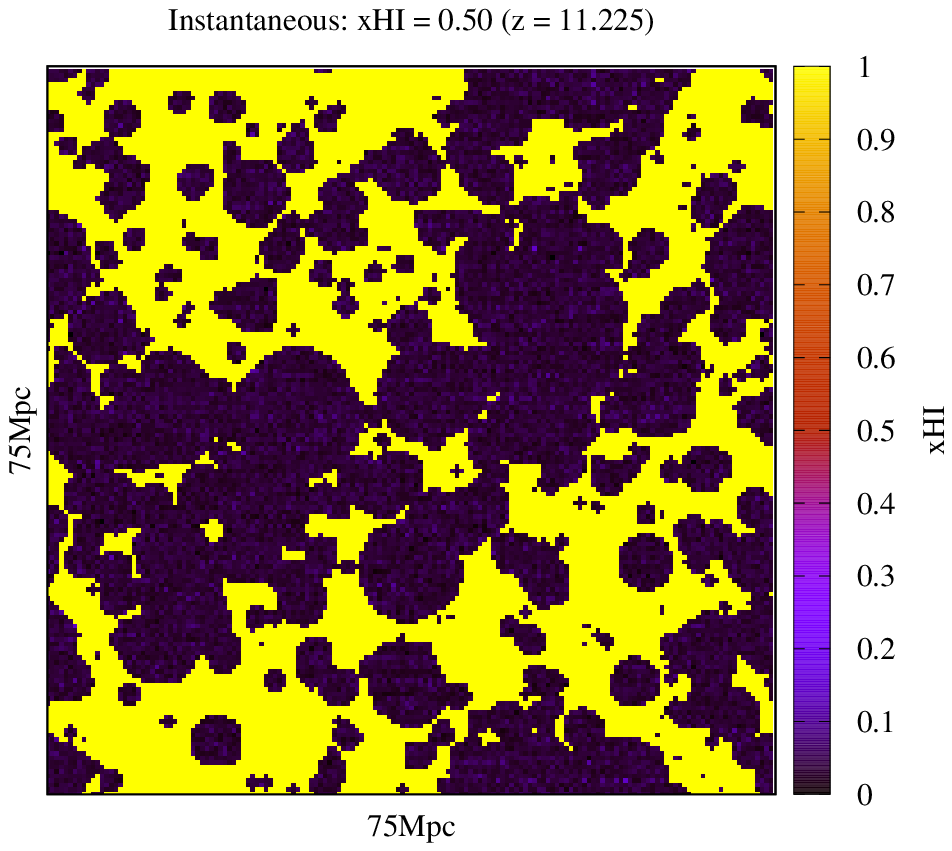}
   \includegraphics[trim = 100 0 120 0, height=6.3cm]{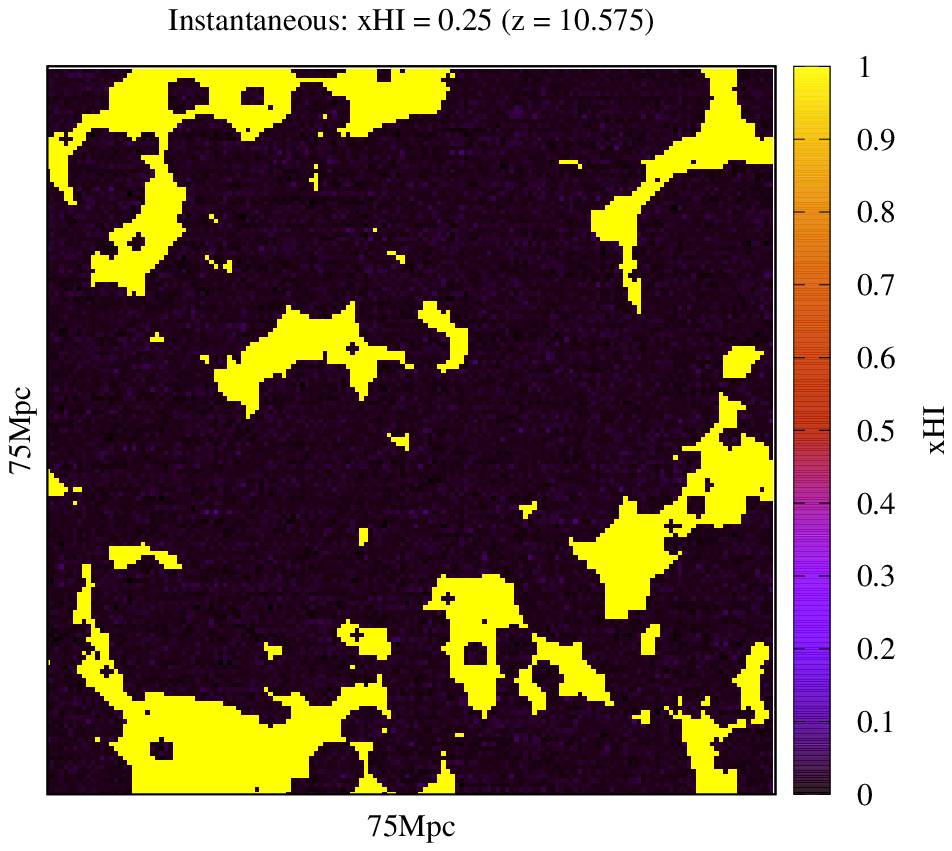}\\
   \includegraphics[trim = 120 0 50 0, height=6.3cm]{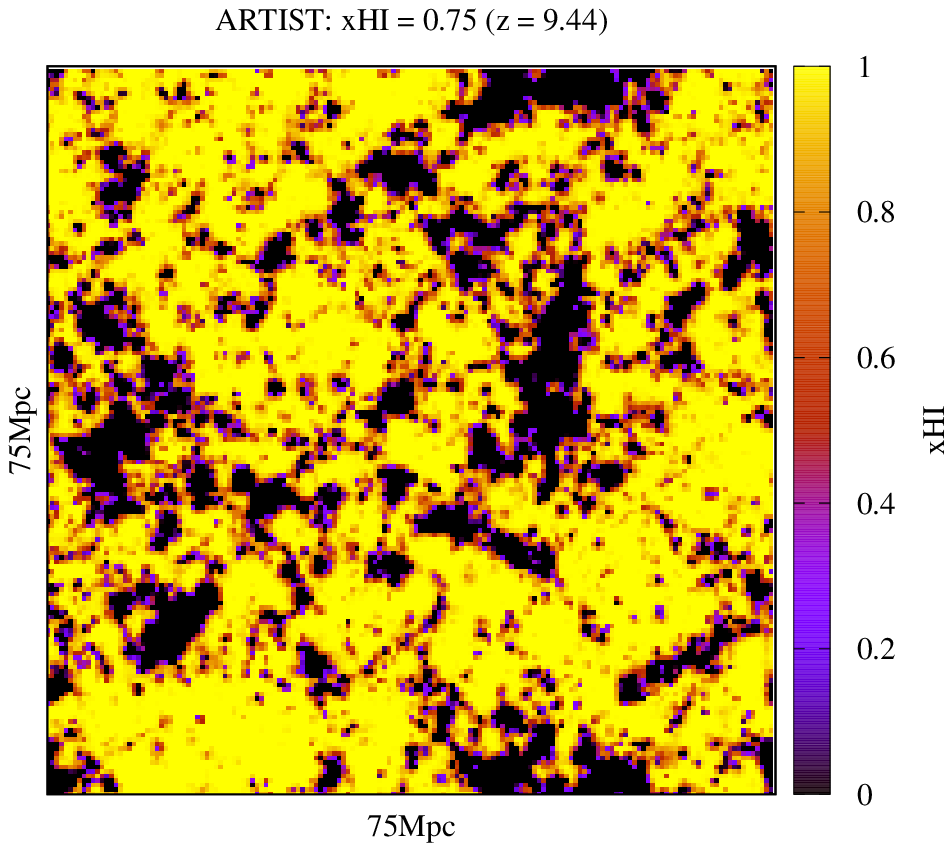}
   \includegraphics[trim = 100 0 50 0, height=6.3cm]{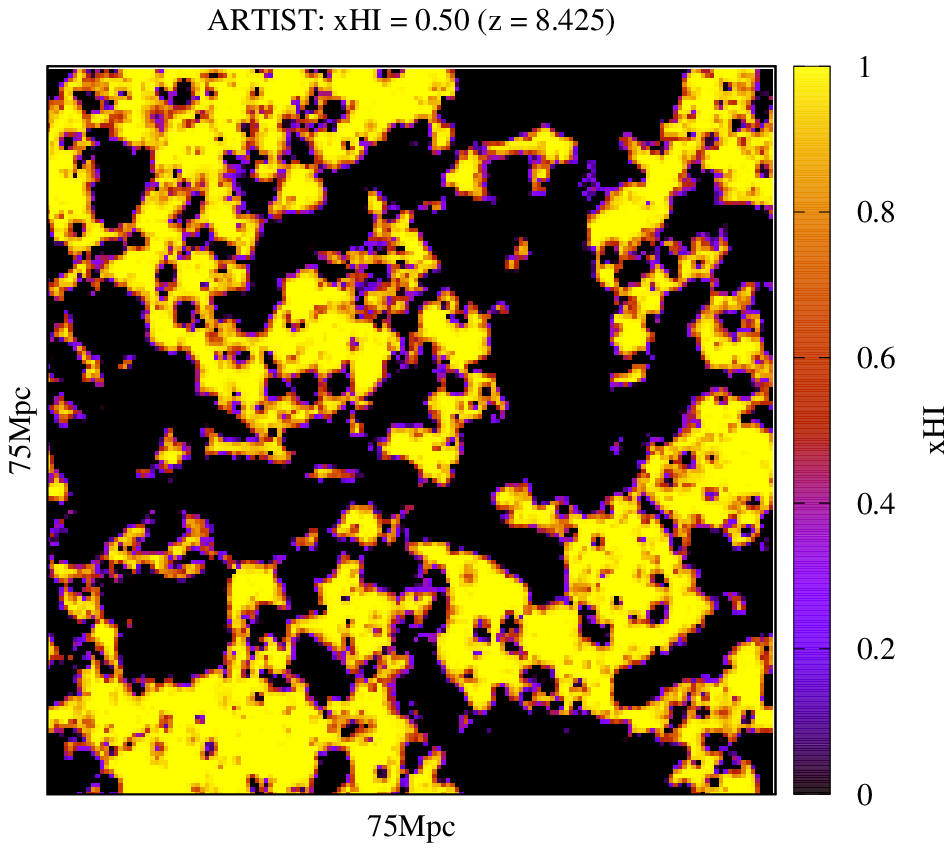}
   \includegraphics[trim = 100 0 120 0, height=6.3cm]{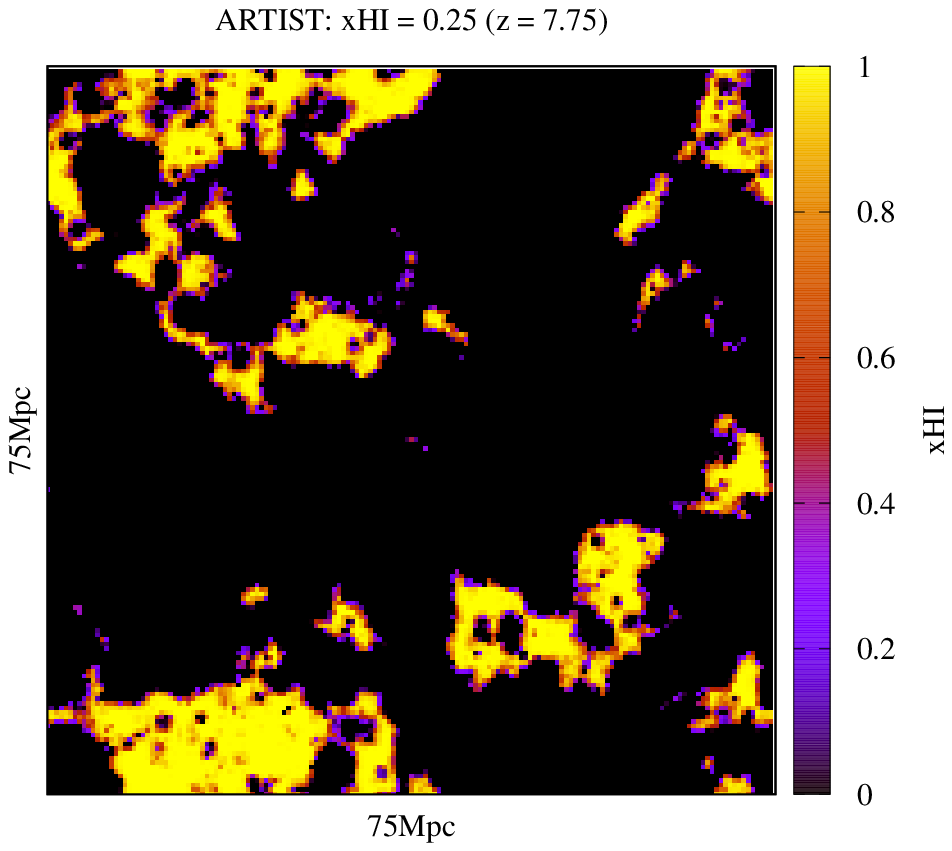}
      
         \caption{Ionisation maps in the 75Mpc box using the instantaneous ESF \textit{(top-row)}, and \ARTIST{} \textit{(bottom-row)} cases. The maps are shown at three values of the global neutral fraction (x$_{\text{HI}}$ = 0.25,0.50 and 0.75), which - due to the different evolution of the neutral gas in the two cases - occur at different redshifts (see Fig. \ref{av_fraction}). This means that although the ionisation morphology shown in \ARTIST{} is quite similar to the one found in the InstESF case, the similar morphologies appear at quite different redshifts.}
         \label{ionisation_maps}
\end{figure*}

In Fig. \ref{ionisation_maps}, we show a slice of our simulation box at three values of the average neutral fraction (x$_{\text{HI}}$ = 0.25, 0.50, and 0.75) for InstESF {(top plot)} and \ARTIST{} {(bottom plot)}. As observed in Fig. \ref{average_ionisation}, the redshifts at which these average neutral fractions occur vary significantly between the different models, as indicated in each plot, but here we focus on morphological characteristics. 

From these maps, we observe that InstESF produces more similar morphologies at the same x$_{\text{HI}}$ values (see Fig. \ref{ionisation_maps}).  However, these occur at very different redshifts (see Fig. \ref{average_ionisation}).  In detail the morphology is more ``blobby", does not follow the filaments quite as well, and again lacks partial ionisation. \ARTIST{} predicts a more complex morphology of the ionisation field than can be captured by the ESF method, presenting a more filamentary-type structure and partial ionisations that cannot be captured by either spherically-averaged ESF approach.

Whereas cells can only be found to be fully ionised or fully neutral by ESF methods (once the size of the ionised region has that of a single cell), \ARTIST's morphology suggests that, during the first stages of the ionisation process, partial ionisation is relevant at the spatial resolution considered for a significant fraction of the volume. Next we consider how these morphological differences reflect in the ionised hydrogen and 21cm power spectra.

\subsection{Power spectrum}
\label{power_spectrum}
In this section we quantify the difference in the ionisation morphology found by the two methods considered by discussing the ionised hydrogen ($P_{\text{xx}}$) and 21cm emission ($P_{\text{21cm}}$) power spectrum for each case. 
These are discussed in their $\Delta^2_{\text{xx}}$ and $\Delta^2_{\text{21cm}}$ form, which are respectively defined as:
\begin{equation}
\label{eqn:Dxx}
\Delta^2_{\text{xx}} \equiv P_{\text{HI}}(k) \frac{k^3}{ 2\uppi^2 x^2_{\text{HI}} }
\end{equation}
and 
\begin{equation}
\label{eqn:D21cm}
\Delta^2_{\text{21cm}} \equiv P_{\text{21cm}}(k) \frac{k^3}{ 2\uppi^2 }
\end{equation}

\begin{figure*}
   \centering
      \includegraphics[trim = 50 0 106 0, height=6cm]{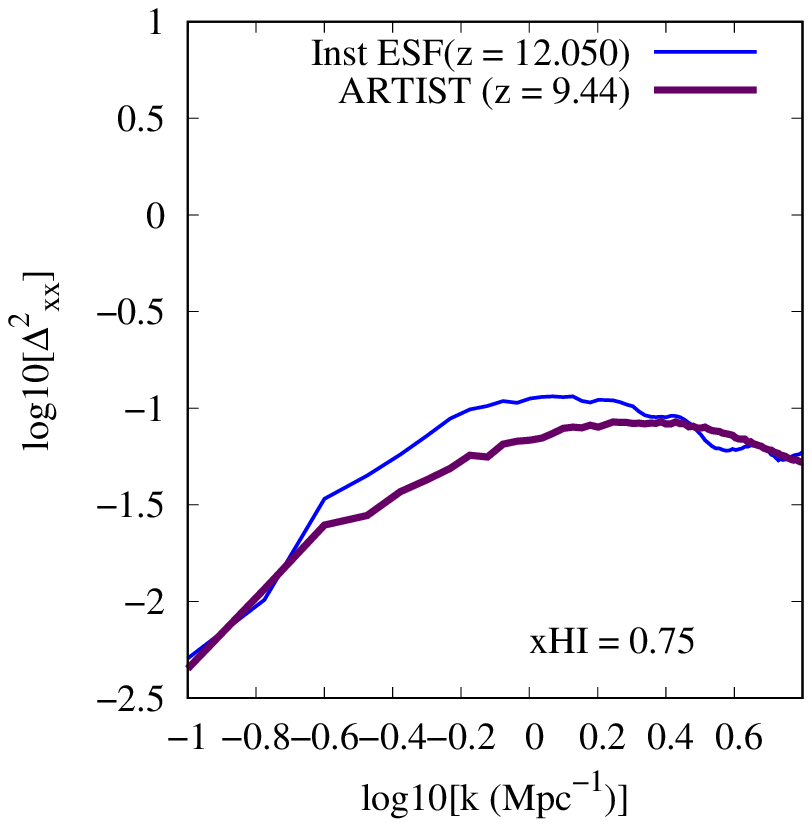}
   \includegraphics[trim = 50 0 125 0, height=6cm]{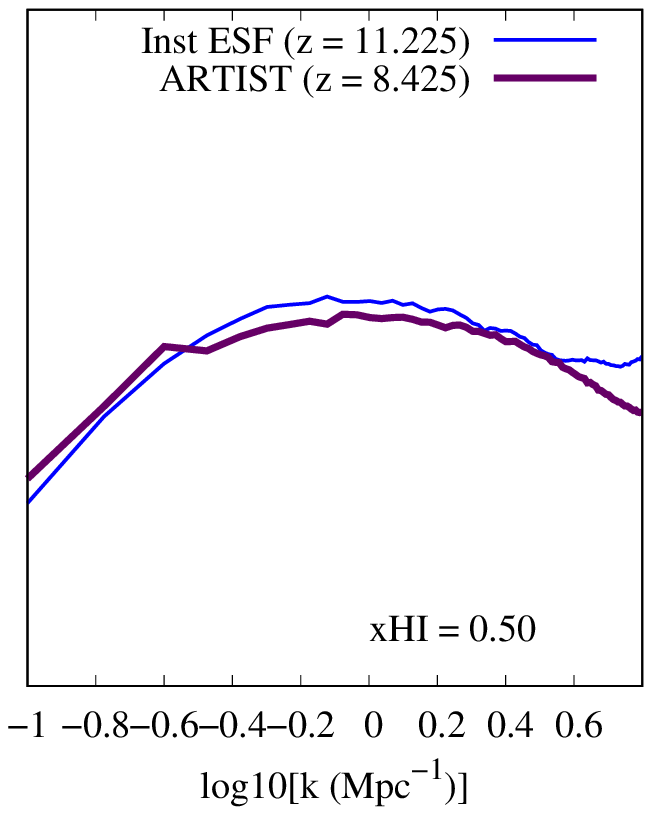}
   \includegraphics[trim = 50 0 100 0, height=6cm]{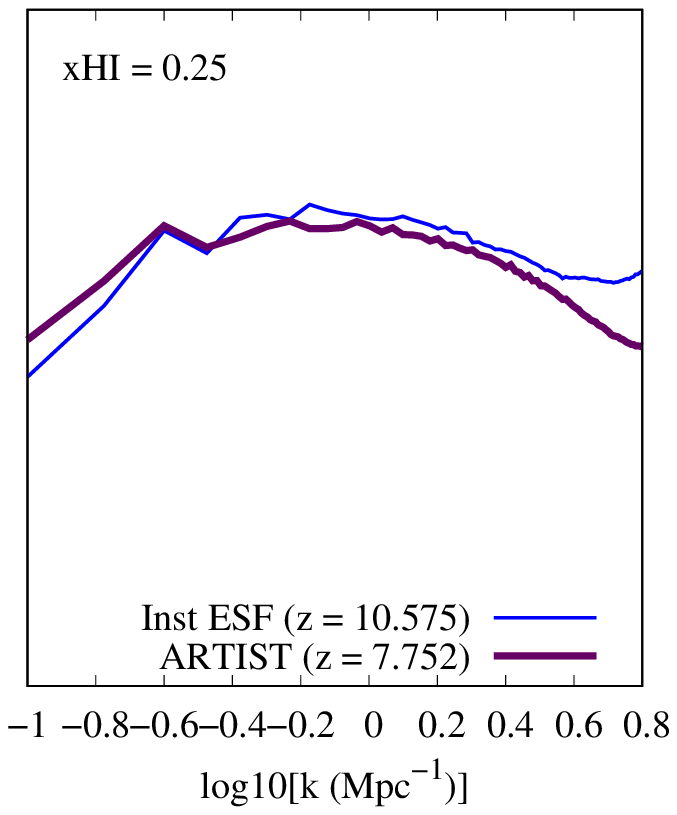}
      \caption{Ionised hydrogen power spectrum $\Delta^2_{\text{xx}}$
      at the average neutral fraction values considered in Fig. \ref{ionisation_maps}, as obtained by \ARTIST{} \textit{(solid purple line)}, and InstESF \textit{(solid blue line)}.             }
           \label{Px}
   \end{figure*}

\begin{figure*}
   \centering
      \includegraphics[trim = 50 0 106 0, height=6cm]{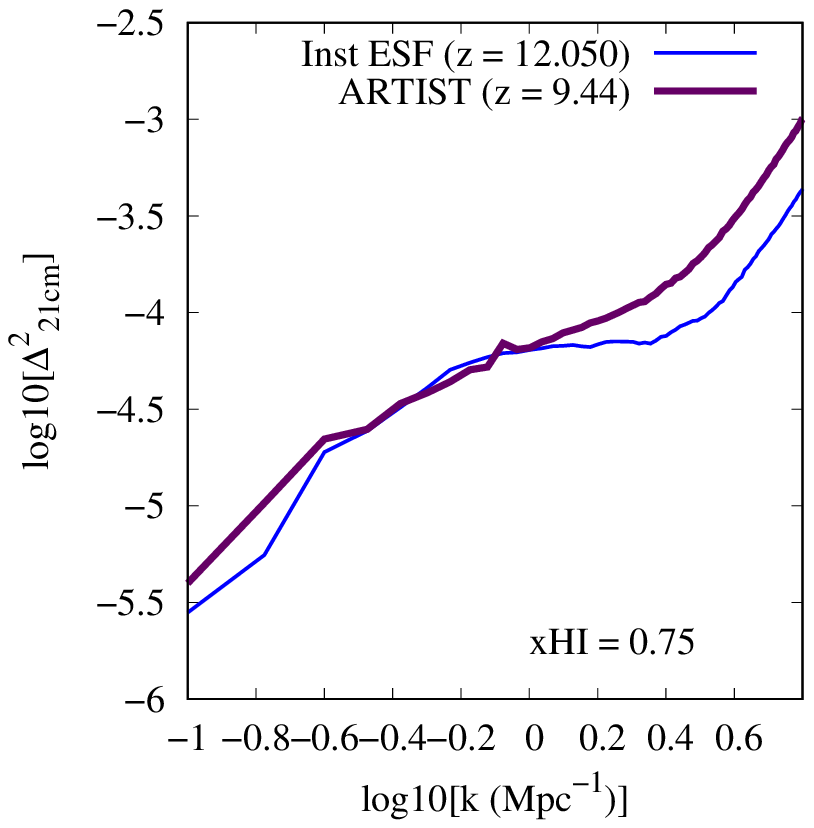}
   \includegraphics[trim = 50 0 125 0, height=6cm]{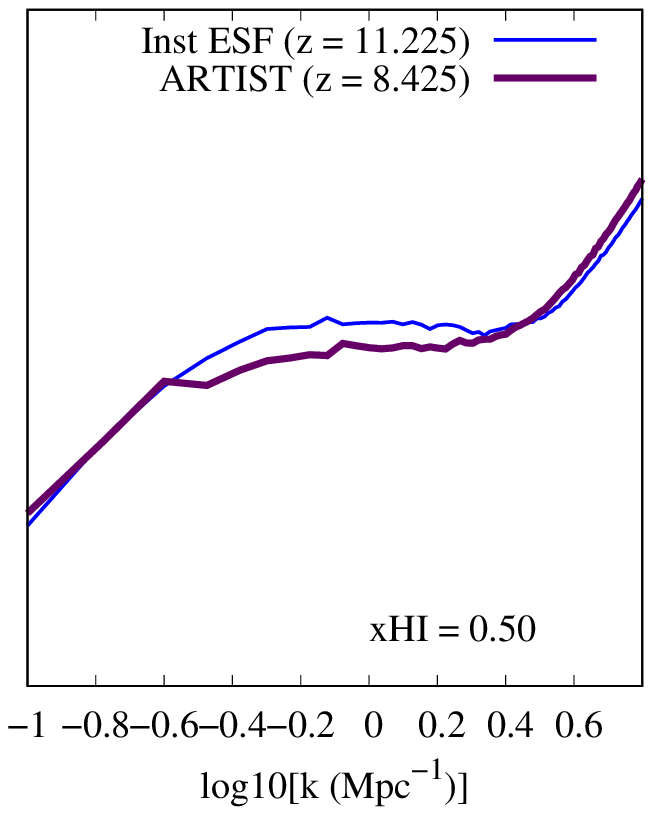}
   \includegraphics[trim = 50 0 100 0, height=6cm]{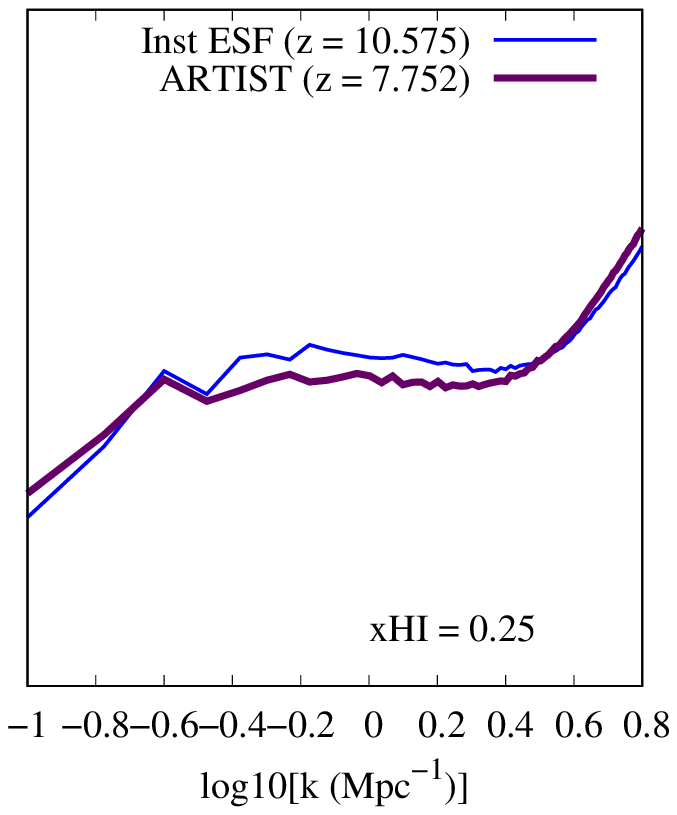}
      \caption{21cm power spectrum $\Delta^2_{21\text{cm}}$ shown in identical fashion to Fig. \ref{Px}.            }
         \label{P21cm}
   \end{figure*}

We show these spectra in Fig. \ref{Px} (for ionised hydrogen) and \ref{P21cm} (for 21cm emission) at the values of global $x_{\text{HI}}$ considered in the previous sections. 
From these figures we observe that the power spectra of InstESF and \ARTIST{} at the same $x_\text{HI}$ is broadly similar, as expected from the similarities in the morphology observed in Fig. \ref{ionisation_maps}. The difference between the two is largest at $x_{\text{HI}} = 0.75$. The reason for the large difference at this $x_{\text{HI}}$ is most likely due to the non-negligible presence of partially ionised, filamentary regions in the \ARTIST{} maps at the beginning of reionisation, which cannot be captured by InstESF. Evidence of the importance of partial ionisations in differentiating the two spectra can be see in the $\Delta^2_{\text{xx}}$ power spectrum evolution of \ARTIST: whereas this increases and then flattens towards smaller scales - as one would expect in the case of non-negligible, partially-ionised regions, InstESF peaks at $k=1$ and then starts decreasing. As previously mentioned, this is a spatial resolution-related effect which ESF is unable to capture at the level of resolution normally considered in its application.

With regards to $\Delta^2_{21\text{cm}}$, in the comparison with InstESF, the partial ionisation in the \ARTIST{} simulation volume leads to the largest differences in the 21cm power spectrum to be observed at the largest and smallest scales, where \ARTIST{} finds more power should be visible at both scales.

\begin{figure}
   \centering
   \includegraphics[trim = 0 0 0 0, width=8cm]{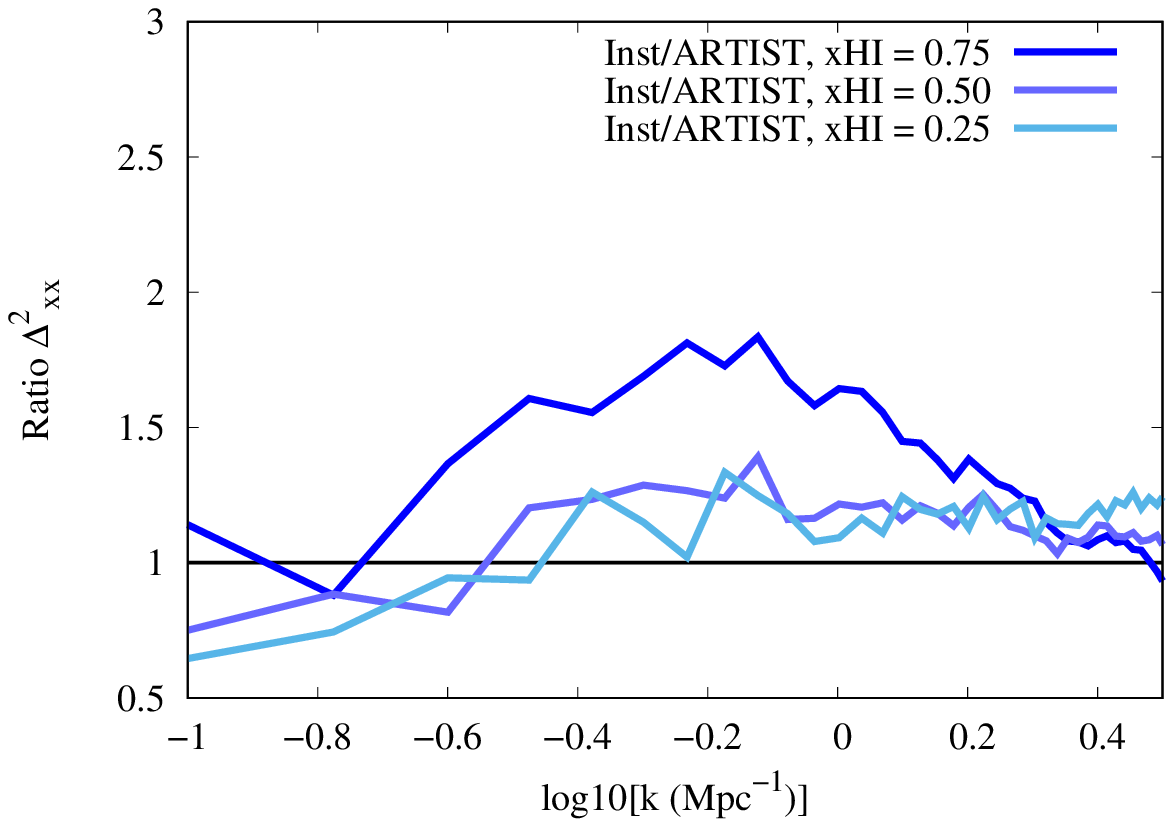} \\
   \includegraphics[trim = 0 0 0 20, width=8cm]{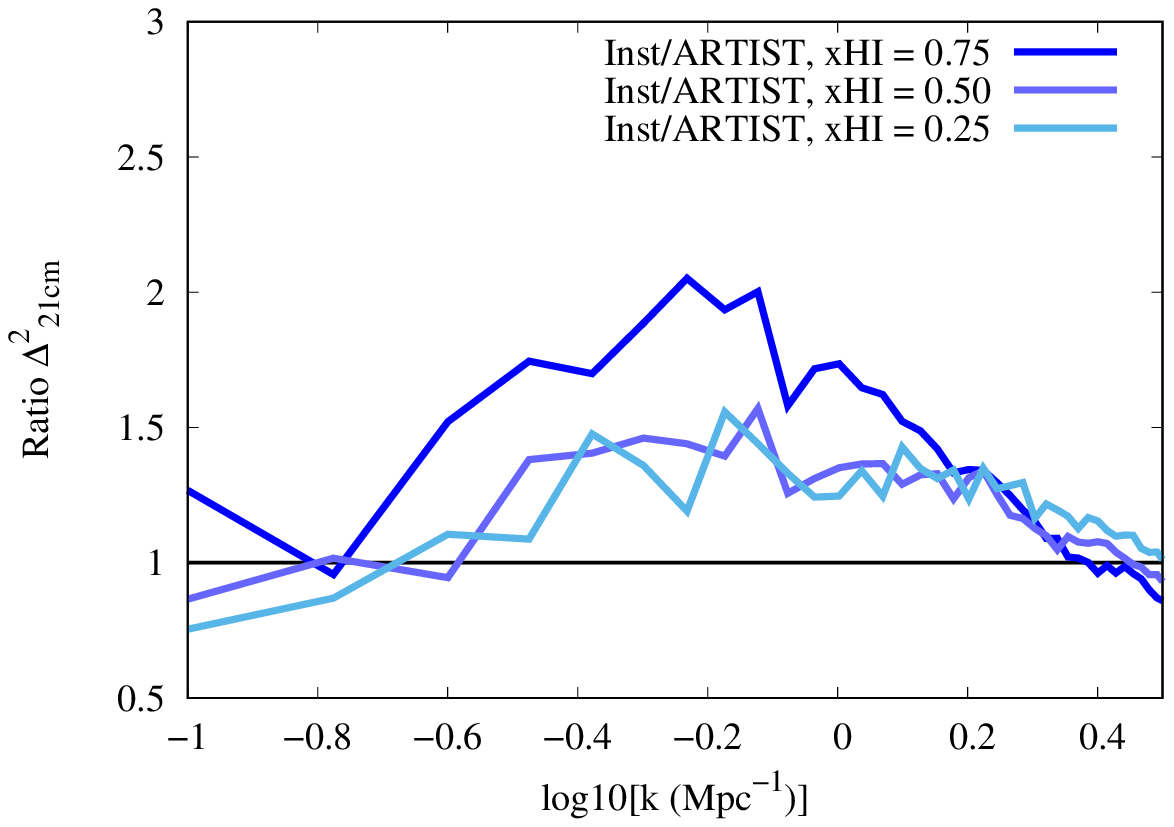}
      \caption{Ratio of \ARTIST's neutral hydrogen \textit{(top plot)} and 21 cm \textit{(bottom plot)} power spectra and those of the Inst \textit{(blue lines)}.
              }
         \label{ratio_power_spectrum}
   \end{figure}

The difference in $\Delta^2$'s amplitudes is more easily quantifiable through the ratio of the power spectra, as shown in Fig. \ref{ratio_power_spectrum}. From this we observe that although the InstESF/\ARTIST{} ratio is indeed very close to 1 at the largest and smallest sizes, it can be up to twice as high for intermediate-size structures, due to the presence of partially ionised structures in \ARTIST.

Our results suggest that the difference in the power spectrum predicted in \ARTIST{} versus the InstESF approach can reach a factor of two. {This is the result of the fact that the InstESF method
tends to wash away signatures of partially-ionised and filamentary structures.}

{We finally note that, despite being able to better match the average neutral fraction evolution, single-cell flagging methods {continue to struggle to conserve the number of photons \citep{Hutter2018}, which is expected to affect their power spectrum predictions \cite{Choudhury2018}}. \ARTIST{} therefore constitutes an improvement in accuracy not only compared to InstESF, but to single-cell flagging methods too.}

\subsection{Summary of ESF vs. \ARTIST{}}
\label{summary_comp_sifast21}
The main findings of this section are as follows:
\begin{itemize}
    \item InstESF produces a somewhat similar morphology to \ARTIST{} at a given global neutral hydrogen fraction. 
    \item The redshift evolution of the average ionisation structure, however, is quite different, with InstESF leading to a much earlier ionisation for a given $f_{\text{esc}}$ {due to issues of photon conservation}. 
    \item The presence of partially ionised and filamentary structures in the \ARTIST{} simulation results in up to a factor of two difference in the amplitude of $\Delta^2_{\text{xx}}$ and $\Delta^2_{\text{21cm}}$. 
\end{itemize}

Overall, \ARTIST{} leads to substantial differences in the EoR evolution when replacing the ESF method currently adopted in {\sc SimFast21}. At a given redshift -- as would be probed by redshifted 21cm observations at a particular frequency -- the differences in the predicted ionised gas and 21cm power spectra can be quite large.  For instance, one would infer a much lower escape fraction from fitting observations to an InstESF-based model vs. \ARTIST. The accuracy of \ARTIST{} over the ESF method is therefore an important improvement for accurately forecasting and interpreting upcoming 21cm observations in order to constrain the nature of the sources that drive reionisation.

\section{Computational performance}
\label{performance}

  \begin{figure}
   \centering
      \includegraphics[trim = 50 0 50 0, height=6cm]{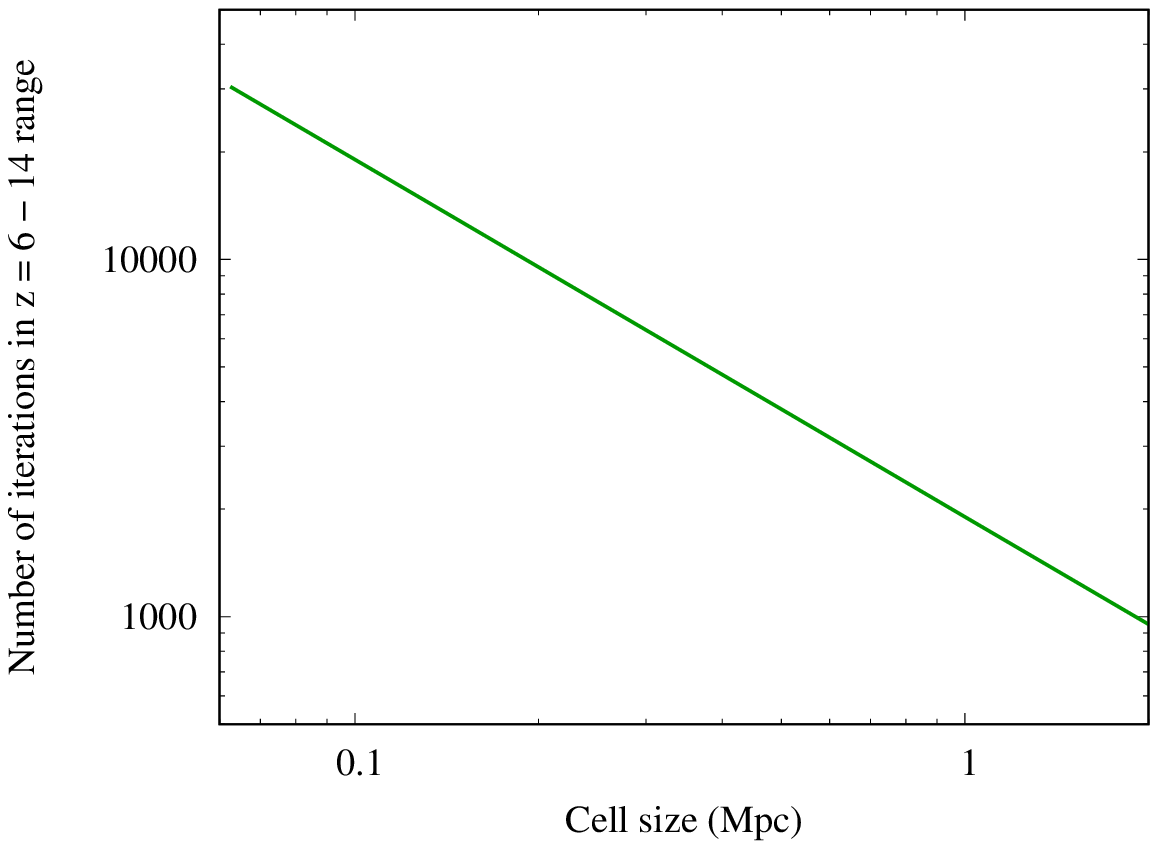}

      \caption{{Number of redshift steps performed in the redshift range 6 -- 14 for difference cell sizes. Notice that this is only dependent on the cell resolution and not on the box size or number of cells independently.} }
           \label{fig:N_vs_dznum}
   \end{figure}  

  \begin{figure}
   \centering
      \includegraphics[trim = 50 0 50 0, height=6cm]{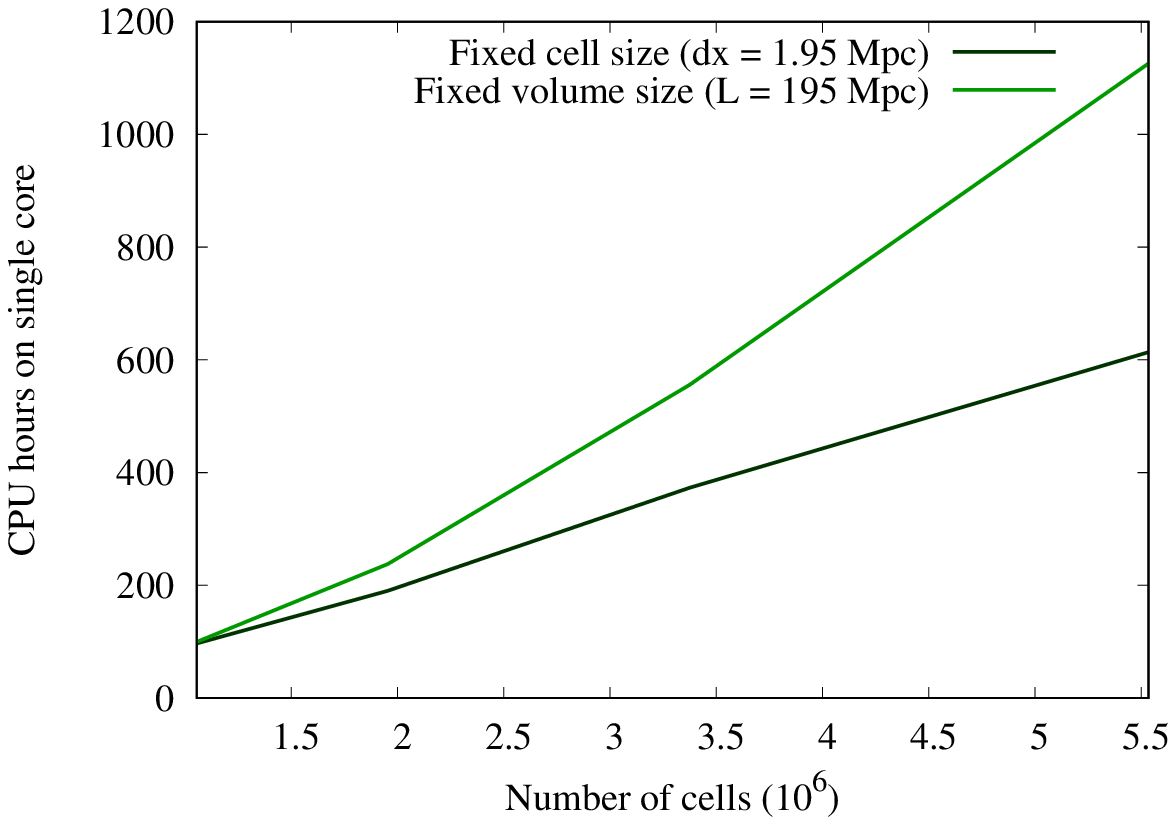}

      \caption{{CPU performance of the code for variable number of cells when fixing the cell size and the simulation volume respectively.} }
           \label{fig:CPU_vs_cells}
   \end{figure}

The advantage of \ARTIST{} in its application to EoR studies is that it can reproduce the results of more complex RT methods (as shown in section \ref{compRT}) while maintaining  modest computational requirements compared to full radiative hydrodynamics codes. {Such requirements can in fact be optimised while safely being able to assess their impact on the accuracy of the simulation}. In this section we quantify these requirements.

\ARTIST{} has been parallelised via OpenMP in order to allow for manageable wall-clock times. Parallelisation of the algorithm is very efficient, as computations of ionisation and excess photons at each cell can be easily computed by different threads across the grid. Provided storage space $\gamma_{l,s,d,i}$ has been pre-allocated to each cell, the same grid parallelisation can also be applied to the redistribution of excess photons across subsequent shell-sections (as described by Eqn. \ref{gen_redistr_d}). 
In our application, however, to limit memory requirements, $\gamma_{l,s,d,i}$ is dynamically allocated only once the $s,d$ shell section of source $l$ has reached cell $i$. Parallel treatment of pre-opened $\gamma_{l,s,d,i}$ storages can still be performed across the grid, provided the cells in which multiple storage spaces have to be opened at the same time step are properly synchronised in the parallel treatment. This allows for the RAM and CPU requirements to be traded off based on individual computational constraints.

{
The computational requirements of \ARTIST{} - just like any ESF-based simulation - scales with the total number of cells considered.} 
{
Unlike ESF-based methods, however, the number of iterations - or redshift steps - that \ARTIST{} performs in the selected redshift range cannot be freely chosen. This is because the time or redshift step that the simulation performs is determined - by the very construction of the algorithm - by the physical size of the cell, as shown in Eqn. \ref{dzstep}. This results in a higher spatial resolution (independently of the number of cells and of the box size), requiring a larger number of redshifts to be sampled in a given redshift range. This is illustrated in Fig. \ref{fig:N_vs_dznum}. }
{
In order to show how \ARTIST{} scales with the number of cells only, therefore, in Fig. \ref{fig:CPU_vs_cells} we show the computational performance of the code when increasing the number of cells but considering the same spatial resolution (shown in the figure by the darker green line), which we achieve by increasing the box size as we increase the number of cells. On the same plot, we also show the total CPU time for different numbers of cells but with a fixed box size, which leads to a higher spatial resolution: because this, as just illustrated, results in a larger number of redshift steps being sampled, the computational cost increases more rapidly. }
{
This further explains why the comparison with the moment-based RT methods considered in section \ref{kristian} is one - even allowing for the fact that \ARTIST{}, unlike the moment-based RT, is mono-frequency - is particularly disadvantageous to \ARTIST{}: the very small box size (12Mpc/h), and hence the even smaller cell size, leads to an extremely large ($\sim 12,700$) number of iterations being performed in the chosen redshift range.} 
{
In the case of the large-scale EoR simulations that \ARTIST{} and ESF methods focus on, however, the number of iterations that will have to be considered will be significantly lower, due to the larger cell sizes considered.}

{In order to perform a more useful performance comparison, therefore, we compare the computational requirements of \ARTIST{} with 21CMMC \citep{Greig2015,Greig2017,Park2019}, a state-of-the-art ESF method optimised for computational efficiency. }
{
In its \cite{Park2019} application, the simulation - which considered a L = 250Mpc box with N=128 cells per side. In this case, the ESF in 21CMMC required a minimum of $\sim 20$ min on a single core, sampling 80 redshifts in this range. The same simulation performed by \ARTIST{} took $12,000$ min on a single core, sampling 160 redshifts. This makes \ARTIST{} 600 times slower in direct comparison, or 300 times slower when directly comparing every redshift iteration. }
      \begin{figure}
   \centering
    \includegraphics[trim = 0 24 0 0, width=8.5cm]{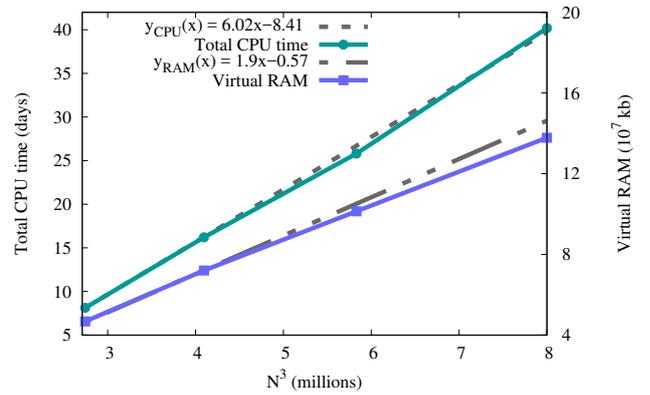}
      \caption{Performance of the simulation in both CPU and RAM as a function of the total number of cells in the simulation. 
              }
         \label{cost_with_N}
   \end{figure}

In Fig. \ref{cost_with_N}, we show the total CPU and RAM required to complete the simulation until the IGM has been completely ionised for the particular case of a 75Mpc box. The total resources requirements are approximately 1000 CPU-hours and 200 GB of RAM for a $200^3$ run down to $z=6$; these are accessible for a single modern workstation.

{\ARTIST{} is therefore able to model large-scale volumes with sufficient dynamic range for upcoming 21cm EoR experiments with modest computational requirements.  MPI parallelisation is also relatively straightforward to access a larger number of cells, but we leave this for future work.}

These manageable requirements, together with the accuracy of the method, {allow \ARTIST{} to produce large-scale EoR simulations of the 21cm power spectrum without replying on un-physical approximations. It is also potentially a way to include RT inexpensively in more sophisticated simulations.}

\section{Discussion and Conclusions}
\label{conclusion}
In this work we have introduced a new method for photon propagation in gridded volumes, the ``Asymmetric Radiative Transfer In Shells Technique'' (\ARTIST), which can successfully reproduce the results of other available radiative transfer (RT) methods (section \ref{compRT}) while reducing the computational costs incurred.
This allows its application to simulations which require large cosmological volumes, such as those studying the Epoch of Reionisation (EoR), to be feasible on physical scales that were previously computationally prohibitive in a RT-based scheme.

\noindent The main features of \ARTIST{} include:
\begin{itemize}
\item the propagation of photons at the speed of light and in a time-dependent fashion;
\item the conservation of  photons in the  simulation, allowing for an accurate estimation of important reionisation parameters {without the need to artificially rescale} $f_{\text{esc}}$;
\item the conservation of the directionality of photons as they propagate away from each source, up to the point when they are incorporated into the background;
\item the accounting of partial  ionisation  in  cells, an  important  feature for  reproducing  the post-reionisation HI measurement;
\item the self-consistent computation of ionisation rates and photon propagation in cells illuminated by multiple sources;
\item the computation of the  ionisation  state  of  each  cell  using  the cell's density, photo-emission, recombination rate and previous ionisation fraction;
\item a variable degrees of angular resolution of the photon propagation around the source, which allows to reproduce shadowing and self-shielding effects with flexible accuracy;
\item the tracking of the  time-dependent  evolution  of  the  radiation field, making it applicable to on-the-fly simulations;
\item thanks to its flexible accuracy and explicit physical assumptions, its application to simulations with different physical and accuracy requirements;
\end{itemize}

In this work, we first demonstrated that the accuracy of our method is consistent with that of other cosmological RTs in benchmark comparison tests considered in \cite{Iliev2006} (see sections \ref{stromgr} and \ref{shadowing}), and showed that \ARTIST{} can nicely reproduce the results of \cite{Finlator2018}'s moment-based RT simulations in a 12Mpc/h box (see section \ref{kristian}) in a tenth of the CPU time. {A comparison between the benchmark RT considered and other cosmological RT methods can be found in \cite{Finlator2009}.}

We then proceeded to apply our RT method to a semi-numerical simulation of the EoR currently adopting an ESF approach, {\sc SimFast21} \citep{Santos2010}, to obtain comparative results. We used \cite{Hassan2017}'s version of the code, and ensured that the only source of difference in the results would be the method of ionisation-fraction estimation, by maintaining the exact same simulation parameters and set-up for all runs. The main findings of our comparison study are as follows:
\begin{itemize}
    \item The use of \ARTIST{} significantly protracts the process of reionisation compared to InstESF. 
    \item  The canonical picture of an EoR morphology characterised by perfectly spherical, fully-ionised regions propagating isotropically from individual sources, is replaced by a more complex picture of the ionised patches, with HII fronts propagating in filament-like structures. \ARTIST{} further suggests that boundary regions in the simulation should be partially ionised to various degrees, more prominently so at higher redshifts. While this effect should disappear at very high resolution, it should appear at spatial resolutions typically considered for EoR applications, and \ARTIST\ is able to track this.
    \item Although the morphology produced by InstESF appears to be qualitative similar, comparisons of $\Delta^2_{\text{xx}}$ and $\Delta^2_{\text{21cm}}$ at fixed $x_{\text{HI}}$ reveal a difference up to a factor of 2 in the power spectra in comparison with \ARTIST. Furthermore, because the redshift evolution of the two is so dissimilar, when comparing the power spectra at fixed redshifts these appear to be completely inconsistent. 

\end{itemize}

{While we considered in particular a whole-sphere flagging ESF method, comparison studies focusing on the relative accuracy of single-cell flagging ESF methods compared to other RT methods have also confirmed discrepancies in the simulated power spectrum, with the likely cause being the lack of photon conservation \cite{Hutter2018}. 
Based on this analysis, we find the differences versus \ARTIST\ and hence the level of inaccuracy, in the ESF approximation to be significant. }\\

This suggests that the ability of ESF-based simulations to make reliable and physically-motivated predictions for future observations of the EoR by experiments such as LOFAR, HERA and the SKA could be compromised, therefore undermining the theoretical interpretation.  \ARTIST, while certainly not perfectly capturing the RT process, {and admittedly being slower than ESF methods by two orders of magnitude, still} presents a compromise in speed and accuracy that reproduces global trends seen in much more expensive calculations relevant for EoR experiments while including more physically well-motivated assumptions than ESF methods.

Among the most severe deficiencies of ESF is the need for the escape fraction in ESF-based analyses to be re-scaled in order to obtain a satisfactory agreement with RT methods. This has recently been addressed by the development of a new ESF method introduced by \cite{Hutter2018}, which achieves reasonable agreement {in the evolution of the average neutral fraction }with RT schemes without requiring an artificial tuning of this parameter.

Issues of photon conservation, however, still remain in all ESF-based methods. As pointed out by the study itself and, more recently, by \cite{Choudhury2018}, the most severe consequence of this is in the redshift evolution of the power spectrum, which \cite{Choudhury2018} attributes to the resolution-dependence of the method. The latter study therefore suggests a new method for post-processing of overlapping ionised regions in order to address the areas in which photon conservation is most problematic.

\ARTIST, on the other hand, as a photon-transport RT algorithm, can consistently account for the number of ionisations taking place in overlapping regions with no need for post- processing.

Furthermore our method, allowing for an asymmetric propagation of photons, better reflects the slow penetration of highly dense regions by the UV photons, rather than averaging higher- and lower-density regions of the ionisation and recombination processes inside spherical volumes.

This increase in the accuracy of the predictions comes at a non-trivial -- but still modest -- computational price.  An \ARTIST{} run that reproduces the neutral fraction evolution from a much more expensive RT-hydro run can be finished within a day or two on a single well-equipped {shared memory cluster}.

\ARTIST{}'s application to large-scale simulations of the EoR thus joins the efforts of numerous new techniques being developed to address the shortcomings of previous ESF methods. {Mutual tests between all models is of course ideal, as it would allow us to better constrain the characteristic limitations affecting each of them. For example, providing a grid-based sampling of large-scale \ARTIST{} simulations could be a useful resource for MCMC applications of ESF methods, maximising the benefit offered by the complementarity of the two methods in their relative advantages.}

As EoR 21cm experiments approach fruition in the coming years, it is crucial that we have theoretical platforms in place to interpret such data in as accurate and robust a way possible.  \ARTIST{} provides a flexible platform upon which to build cosmological EoR models where computational efficiency is crucial but the accuracy can be relaxed as needed. An extension to multi-frequency photon propagation required to model Helium reionisation is straightforward, and the source and recombination terms can be easily extended to include AGN or Population III star contributions.  Given its efficiency, it is even potentially feasible to include \ARTIST{} on the fly in full hydrodynamic simulations, thereby self-consistently generating an ionising background during the EoR and beyond from the dynamically modeled galaxy population.  These extensions are in progress.

\section*{Acknowledgements}
Our simulations were run on the following computing facilities: the Pumba Astrophysics Computing Cluster, hosted at the University of the Western Cape (UWC), which was generously funded by UWC’s Office of the Deputy Vice Chancellor; and the ilify cloud computing facility, hosted at the University of Cape Town (UCT) as a partnership between UCT, UWC, the University of Stellenbosch, Sol Plaatje University, the Cape Peninsula University of Technology, and the South African Radio Astronomy Observatory. The ilifu facility is supported by contributions from the
Inter-University Institute for Data Intensive Astronomy (IDIA - a partnership between UCT, UWC, and the University of Pretoria), the Computational Biology division at UCT, and the Data Intensive Research Initiative of South Africa (DIRISA). The Technicolor
simulations used the Extreme Science and Engineering Discovery
Environment (XSEDE), which is supported by National Science
Foundation grant number ACI-1548562. RD acknowledges support
from the Wolfson Research Merit Award programme of the UK
Royal Society. We thank the anonymous referee for the very useful
feedback on our draft, which helped us improve our paper.

\bibliographystyle{mnras}
\bibliography{references}

\begin{thebibliography}{}
\makeatletter
\relax
\def\mn@urlcharsother{\let\do\@makeother \do\$\do\&\do\#\do\^\do\_\do\%\do\~}
\def\mn@doi{\begingroup\mn@urlcharsother \@ifnextchar [ {\mn@doi@}
  {\mn@doi@[]}}
\def\mn@doi@[#1]#2{\def\@tempa{#1}\ifx\@tempa\@empty \href
  {http://dx.doi.org/#2} {doi:#2}\else \href {http://dx.doi.org/#2} {#1}\fi
  \endgroup}
\def\mn@eprint#1#2{\mn@eprint@#1:#2::\@nil}
\def\mn@eprint@arXiv#1{\href {http://arxiv.org/abs/#1} {{\tt arXiv:#1}}}
\def\mn@eprint@dblp#1{\href {http://dblp.uni-trier.de/rec/bibtex/#1.xml}
  {dblp:#1}}
\def\mn@eprint@#1:#2:#3:#4\@nil{\def\@tempa {#1}\def\@tempb {#2}\def\@tempc
  {#3}\ifx \@tempc \@empty \let \@tempc \@tempb \let \@tempb \@tempa \fi \ifx
  \@tempb \@empty \def\@tempb {arXiv}\fi \@ifundefined
  {mn@eprint@\@tempb}{\@tempb:\@tempc}{\expandafter \expandafter \csname
  mn@eprint@\@tempb\endcsname \expandafter{\@tempc}}}

\bibitem[\protect\citeauthoryear{{Altay}, {Croft}  \& {Pelupessy}}{{Altay}
  et~al.}{2008}]{Altay2008}
{Altay} G.,  {Croft} R.~A.~C.,   {Pelupessy} I.,  2008, \mn@doi [\mnras]
  {10.1111/j.1365-2966.2008.13212.x}, \href
  {http://adsabs.harvard.edu/abs/2008MNRAS.386.1931A} {386, 1931}

\bibitem[\protect\citeauthoryear{{Alvarez}, {Busha}, {Abel}  \&
  {Wechsler}}{{Alvarez} et~al.}{2009}]{Alvarez2009}
{Alvarez} M.~A.,  {Busha} M.,  {Abel} T.,   {Wechsler} R.~H.,  2009, \mn@doi
  [\apjl] {10.1088/0004-637X/703/2/L167}, \href
  {http://adsabs.harvard.edu/abs/2009ApJ...703L.167A} {703, L167}

\bibitem[\protect\citeauthoryear{{Aubert} \& {Teyssier}}{{Aubert} \&
  {Teyssier}}{2008}]{Aubert2008}
{Aubert} D.,  {Teyssier} R.,  2008, \mn@doi [\mnras]
  {10.1111/j.1365-2966.2008.13223.x}, \href
  {http://adsabs.harvard.edu/abs/2008MNRAS.387..295A} {387, 295}

\bibitem[\protect\citeauthoryear{{Barkana} \& {Loeb}}{{Barkana} \&
  {Loeb}}{2001}]{Barkana2001}
{Barkana} R.,  {Loeb} A.,  2001, \mn@doi [\physrep]
  {10.1016/S0370-1573(01)00019-9}, \href
  {http://adsabs.harvard.edu/abs/2001PhR...349..125B} {349, 125}

\bibitem[\protect\citeauthoryear{{Bauer}, {Springel}, {Vogelsberger}, {Genel},
  {Torrey}, {Sijacki}, {Nelson}  \& {Hernquist}}{{Bauer}
  et~al.}{2015}]{Bauer2015}
{Bauer} A.,  {Springel} V.,  {Vogelsberger} M.,  {Genel} S.,  {Torrey} P.,
  {Sijacki} D.,  {Nelson} D.,   {Hernquist} L.,  2015, \mn@doi [\mnras]
  {10.1093/mnras/stv1893}, \href
  {http://adsabs.harvard.edu/abs/2015MNRAS.453.3593B} {453, 3593}

\bibitem[\protect\citeauthoryear{{Bond}, {Cole}, {Efstathiou}  \&
  {Kaiser}}{{Bond} et~al.}{1991}]{Bond1991}
{Bond} J.~R.,  {Cole} S.,  {Efstathiou} G.,   {Kaiser} N.,  1991, \mn@doi
  [\apj] {10.1086/170520}, \href
  {http://adsabs.harvard.edu/abs/1991ApJ...379..440B} {379, 440}

\bibitem[\protect\citeauthoryear{{Bowman} et~al.,}{{Bowman}
  et~al.}{2013}]{Bowman2013}
{Bowman} J.~D.,  et~al., 2013, \mn@doi [\pasa] {10.1017/pas.2013.009}, \href
  {http://adsabs.harvard.edu/abs/2013PASA...30...31B} {30, e031}

\bibitem[\protect\citeauthoryear{{Choudhury} \& {Paranjape}}{{Choudhury} \&
  {Paranjape}}{2018}]{Choudhury2018}
{Choudhury} T.~R.,  {Paranjape} A.,  2018, \mn@doi [\mnras]
  {10.1093/mnras/sty2551}, \href
  {https://ui.adsabs.harvard.edu/abs/2018MNRAS.481.3821C} {481, 3821}

\bibitem[\protect\citeauthoryear{{Choudhury}, {Haehnelt}  \&
  {Regan}}{{Choudhury} et~al.}{2009}]{Choudhury2009}
{Choudhury} T.~R.,  {Haehnelt} M.~G.,   {Regan} J.,  2009, \mn@doi [\mnras]
  {10.1111/j.1365-2966.2008.14383.x}, \href
  {http://adsabs.harvard.edu/abs/2009MNRAS.394..960C} {394, 960}

\bibitem[\protect\citeauthoryear{{Ciardi}, {Ferrara}  \& {White}}{{Ciardi}
  et~al.}{2003}]{Ciardi2003}
{Ciardi} B.,  {Ferrara} A.,   {White} S.~D.~M.,  2003, \mn@doi [\mnras]
  {10.1046/j.1365-8711.2003.06976.x}, \href
  {http://adsabs.harvard.edu/abs/2003MNRAS.344L...7C} {344, L7}

\bibitem[\protect\citeauthoryear{{D'Aloisio}, {McQuinn}, {Maupin}, {Davies},
  {Trac}, {Fuller}  \& {Upton Sanderbeck}}{{D'Aloisio}
  et~al.}{2019}]{daloisio2019}
{D'Aloisio} A.,  {McQuinn} M.,  {Maupin} O.,  {Davies} F.~B.,  {Trac} H.,
  {Fuller} S.,   {Upton Sanderbeck} P.~R.,  2019, \mn@doi [The Astrophysical
  Journal] {10.3847/1538-4357/ab0d83}, \href
  {https://ui.adsabs.harvard.edu/abs/2019ApJ...874..154D} {874, 154}

\bibitem[\protect\citeauthoryear{{Dav{\'e}}, {Katz}, {Oppenheimer}, {Kollmeier}
   \& {Weinberg}}{{Dav{\'e}} et~al.}{2013}]{Dave2013}
{Dav{\'e}} R.,  {Katz} N.,  {Oppenheimer} B.~D.,  {Kollmeier} J.~A.,
  {Weinberg} D.~H.,  2013, \mn@doi [\mnras] {10.1093/mnras/stt1274}, \href
  {http://adsabs.harvard.edu/abs/2013MNRAS.434.2645D} {434, 2645}

\bibitem[\protect\citeauthoryear{{Dayal} \& {Ferrara}}{{Dayal} \&
  {Ferrara}}{2018}]{Dayal2018}
{Dayal} P.,  {Ferrara} A.,  2018, \mn@doi [\physrep]
  {10.1016/j.physrep.2018.10.002}, \href
  {https://ui.adsabs.harvard.edu/abs/2018PhR...780....1D} {780, 1}

\bibitem[\protect\citeauthoryear{{DeBoer} et~al.,}{{DeBoer}
  et~al.}{2017}]{DeBoer2017}
{DeBoer} D.~R.,  et~al., 2017, \mn@doi [\pasp]
  {10.1088/1538-3873/129/974/045001}, \href
  {http://adsabs.harvard.edu/abs/2017PASP..129d5001D} {129, 045001}

\bibitem[\protect\citeauthoryear{{Fan}, {Carilli}  \& {Keating}}{{Fan}
  et~al.}{2006}]{Fan2006}
{Fan} X.,  {Carilli} C.~L.,   {Keating} B.,  2006, \mn@doi [\araa]
  {10.1146/annurev.astro.44.051905.092514}, \href
  {http://adsabs.harvard.edu/abs/2006ARA%26A..44..415F} {44, 415}

\bibitem[\protect\citeauthoryear{{Finlator}, {{\"O}zel}  \&
  {Dav{\'e}}}{{Finlator} et~al.}{2009}]{Finlator2009}
{Finlator} K.,  {{\"O}zel} F.,   {Dav{\'e}} R.,  2009, \mn@doi [\mnras]
  {10.1111/j.1365-2966.2008.14190.x}, \href
  {http://adsabs.harvard.edu/abs/2009MNRAS.393.1090F} {393, 1090}

\bibitem[\protect\citeauthoryear{{Finlator}, {Mu{\~n}oz}, {Oppenheimer}, {Oh},
  {{\"O}zel}  \& {Dav{\'e}}}{{Finlator} et~al.}{2013}]{Finlator2013}
{Finlator} K.,  {Mu{\~n}oz} J.~A.,  {Oppenheimer} B.~D.,  {Oh} S.~P.,
  {{\"O}zel} F.,   {Dav{\'e}} R.,  2013, \mn@doi [\mnras]
  {10.1093/mnras/stt1697}, \href
  {http://adsabs.harvard.edu/abs/2013MNRAS.436.1818F} {436, 1818}

\bibitem[\protect\citeauthoryear{{Finlator}, {Thompson}, {Huang}, {Dav{\'e}},
  {Zackrisson}  \& {Oppenheimer}}{{Finlator} et~al.}{2015}]{Finlator2015}
{Finlator} K.,  {Thompson} R.,  {Huang} S.,  {Dav{\'e}} R.,  {Zackrisson} E.,
  {Oppenheimer} B.~D.,  2015, \mn@doi [\mnras] {10.1093/mnras/stu2668}, \href
  {http://adsabs.harvard.edu/abs/2015MNRAS.447.2526F} {447, 2526}

\bibitem[\protect\citeauthoryear{{Finlator} et~al.,}{{Finlator}
  et~al.}{2017}]{Finlator2017}
{Finlator} K.,  et~al., 2017, \mn@doi [\mnras] {10.1093/mnras/stw2433}, \href
  {http://adsabs.harvard.edu/abs/2017MNRAS.464.1633F} {464, 1633}

\bibitem[\protect\citeauthoryear{{Finlator}, {Keating}, {Oppenheimer},
  {Dav{\'e}}  \& {Zackrisson}}{{Finlator} et~al.}{2018}]{Finlator2018}
{Finlator} K.,  {Keating} L.,  {Oppenheimer} B.~D.,  {Dav{\'e}} R.,
  {Zackrisson} E.,  2018, preprint, \href
  {http://adsabs.harvard.edu/abs/2018arXiv180500099F} {} (\mn@eprint {arXiv}
  {1805.00099})

\bibitem[\protect\citeauthoryear{{Furlanetto} \& {Oh}}{{Furlanetto} \&
  {Oh}}{2005}]{Furlanetto2005}
{Furlanetto} S.~R.,  {Oh} S.~P.,  2005, \mn@doi [\mnras]
  {10.1111/j.1365-2966.2005.09505.x}, \href
  {http://adsabs.harvard.edu/abs/2005MNRAS.363.1031F} {363, 1031}

\bibitem[\protect\citeauthoryear{{Geil} \& {Wyithe}}{{Geil} \&
  {Wyithe}}{2008}]{Geil2008}
{Geil} P.~M.,  {Wyithe} J.~S.~B.,  2008, \mn@doi [\mnras]
  {10.1111/j.1365-2966.2008.13159.x}, \href
  {http://adsabs.harvard.edu/abs/2008MNRAS.386.1683G} {386, 1683}

\bibitem[\protect\citeauthoryear{{Gnedin}}{{Gnedin}}{2000}]{Gnedin2000}
{Gnedin} N.~Y.,  2000, \mn@doi [\apj] {10.1086/317042}, \href
  {http://adsabs.harvard.edu/abs/2000ApJ...542..535G} {542, 535}

\bibitem[\protect\citeauthoryear{{Gnedin}}{{Gnedin}}{2014}]{Gnedin2014}
{Gnedin} N.~Y.,  2014, \mn@doi [\apj] {10.1088/0004-637X/793/1/29}, \href
  {http://adsabs.harvard.edu/abs/2014ApJ...793...29G} {793, 29}

\bibitem[\protect\citeauthoryear{{Graziani}, {Maselli}  \& {Ciardi}}{{Graziani}
  et~al.}{2013}]{Graziani2013}
{Graziani} L.,  {Maselli} A.,   {Ciardi} B.,  2013, \mn@doi [\mnras]
  {10.1093/mnras/stt206}, \href
  {http://adsabs.harvard.edu/abs/2013MNRAS.431..722G} {431, 722}

\bibitem[\protect\citeauthoryear{{Greig} \& {Mesinger}}{{Greig} \&
  {Mesinger}}{2015}]{Greig2015}
{Greig} B.,  {Mesinger} A.,  2015, \mn@doi [\mnras] {10.1093/mnras/stv571},
  \href {https://ui.adsabs.harvard.edu/abs/2015MNRAS.449.4246G} {449, 4246}

\bibitem[\protect\citeauthoryear{{Greig} \& {Mesinger}}{{Greig} \&
  {Mesinger}}{2017}]{Greig2017}
{Greig} B.,  {Mesinger} A.,  2017, \mn@doi [\mnras] {10.1093/mnras/stx2118},
  \href {http://adsabs.harvard.edu/abs/2017MNRAS.472.2651G} {472, 2651}

\bibitem[\protect\citeauthoryear{{Hassan}, {Dav{\'e}}, {Finlator}  \&
  {Santos}}{{Hassan} et~al.}{2016}]{Hassan2016}
{Hassan} S.,  {Dav{\'e}} R.,  {Finlator} K.,   {Santos} M.~G.,  2016, \mn@doi
  [\mnras] {10.1093/mnras/stv3001}, \href
  {http://adsabs.harvard.edu/abs/2016MNRAS.457.1550H} {457, 1550}

\bibitem[\protect\citeauthoryear{{Hassan}, {Dav{\'e}}, {Finlator}  \&
  {Santos}}{{Hassan} et~al.}{2017}]{Hassan2017}
{Hassan} S.,  {Dav{\'e}} R.,  {Finlator} K.,   {Santos} M.~G.,  2017, \mn@doi
  [\mnras] {10.1093/mnras/stx420}, \href
  {http://adsabs.harvard.edu/abs/2017MNRAS.468..122H} {468, 122}

\bibitem[\protect\citeauthoryear{{Hutter}}{{Hutter}}{2018}]{Hutter2018}
{Hutter} A.,  2018, \mn@doi [\mnras] {10.1093/mnras/sty683}, \href
  {http://adsabs.harvard.edu/abs/2018MNRAS.477.1549H} {477, 1549}

\bibitem[\protect\citeauthoryear{{Iliev} et~al.,}{{Iliev}
  et~al.}{2006}]{Iliev2006}
{Iliev} I.~T.,  et~al., 2006, \mn@doi [\mnras]
  {10.1111/j.1365-2966.2006.10775.x}, \href
  {http://adsabs.harvard.edu/abs/2006MNRAS.371.1057I} {371, 1057}

\bibitem[\protect\citeauthoryear{{Iliev}, {Mellema}, {Ahn}, {Shapiro}, {Mao}
  \& {Pen}}{{Iliev} et~al.}{2014}]{Iliev2014}
{Iliev} I.~T.,  {Mellema} G.,  {Ahn} K.,  {Shapiro} P.~R.,  {Mao} Y.,   {Pen}
  U.-L.,  2014, \mn@doi [\mnras] {10.1093/mnras/stt2497}, \href
  {http://adsabs.harvard.edu/abs/2014MNRAS.439..725I} {439, 725}

\bibitem[\protect\citeauthoryear{{Iliev}, {Santos}, {Mesinger}, {Majumdar}  \&
  {Mellema}}{{Iliev} et~al.}{2015}]{Iliev2015}
{Iliev} I.,  {Santos} M.,  {Mesinger} A.,  {Majumdar} S.,   {Mellema} G.,
  2015, Advancing Astrophysics with the Square Kilometre Array (AASKA14), \href
  {http://adsabs.harvard.edu/abs/2015aska.confE...7I} {p.~7}

\bibitem[\protect\citeauthoryear{{Katz}, {Kimm}, {Sijacki}  \&
  {Haehnelt}}{{Katz} et~al.}{2017}]{Katz2017}
{Katz} H.,  {Kimm} T.,  {Sijacki} D.,   {Haehnelt} M.~G.,  2017, \mn@doi
  [\mnras] {10.1093/mnras/stx608}, \href
  {http://adsabs.harvard.edu/abs/2017MNRAS.468.4831K} {468, 4831}

\bibitem[\protect\citeauthoryear{{Loeb} \& {Barkana}}{{Loeb} \&
  {Barkana}}{2001}]{Loeb2001}
{Loeb} A.,  {Barkana} R.,  2001, \mn@doi [\araa]
  {10.1146/annurev.astro.39.1.19}, \href
  {http://adsabs.harvard.edu/abs/2001ARA%26A..39...19L} {39, 19}

\bibitem[\protect\citeauthoryear{{McQuinn}, {Furlanetto}, {Hernquist}, {Zahn}
  \& {Zaldarriaga}}{{McQuinn} et~al.}{2005}]{McQuinn2005}
{McQuinn} M.,  {Furlanetto} S.~R.,  {Hernquist} L.,  {Zahn} O.,   {Zaldarriaga}
  M.,  2005, \mn@doi [\apj] {10.1086/432049}, \href
  {http://adsabs.harvard.edu/abs/2005ApJ...630..643M} {630, 643}

\bibitem[\protect\citeauthoryear{{McQuinn}, {Lidz}, {Zahn}, {Dutta},
  {Hernquist}  \& {Zaldarriaga}}{{McQuinn} et~al.}{2007}]{McQuinn2007}
{McQuinn} M.,  {Lidz} A.,  {Zahn} O.,  {Dutta} S.,  {Hernquist} L.,
  {Zaldarriaga} M.,  2007, \mn@doi [\mnras] {10.1111/j.1365-2966.2007.11489.x},
  \href {http://adsabs.harvard.edu/abs/2007MNRAS.377.1043M} {377, 1043}

\bibitem[\protect\citeauthoryear{{Mellema}, {Iliev}, {Alvarez}  \&
  {Shapiro}}{{Mellema} et~al.}{2006}]{Mellema2006}
{Mellema} G.,  {Iliev} I.~T.,  {Alvarez} M.~A.,   {Shapiro} P.~R.,  2006,
  \mn@doi [\na] {10.1016/j.newast.2005.09.004}, \href
  {http://adsabs.harvard.edu/abs/2006NewA...11..374M} {11, 374}

\bibitem[\protect\citeauthoryear{{Mellema}, {Koopmans}, {Shukla}, {Datta},
  {Mesinger}  \& {Majumdar}}{{Mellema} et~al.}{2015}]{Mellema2015}
{Mellema} G.,  {Koopmans} L.,  {Shukla} H.,  {Datta} K.~K.,  {Mesinger} A.,
  {Majumdar} S.,  2015, Advancing Astrophysics with the Square Kilometre Array
  (AASKA14), \href {http://adsabs.harvard.edu/abs/2015aska.confE..10M} {p.~10}

\bibitem[\protect\citeauthoryear{{Mesinger} \& {Furlanetto}}{{Mesinger} \&
  {Furlanetto}}{2007}]{Mesinger2007}
{Mesinger} A.,  {Furlanetto} S.,  2007, \mn@doi [\apj] {10.1086/521806}, \href
  {http://adsabs.harvard.edu/abs/2007ApJ...669..663M} {669, 663}

\bibitem[\protect\citeauthoryear{{Mesinger}, {Furlanetto}  \& {Cen}}{{Mesinger}
  et~al.}{2011}]{Mesinger2011}
{Mesinger} A.,  {Furlanetto} S.,   {Cen} R.,  2011, \mn@doi [\mnras]
  {10.1111/j.1365-2966.2010.17731.x}, \href
  {http://adsabs.harvard.edu/abs/2011MNRAS.411..955M} {411, 955}

\bibitem[\protect\citeauthoryear{{Noh} \& {McQuinn}}{{Noh} \&
  {McQuinn}}{2014}]{Noh2014}
{Noh} Y.,  {McQuinn} M.,  2014, \mn@doi [\mnras] {10.1093/mnras/stu1412}, \href
  {http://adsabs.harvard.edu/abs/2014MNRAS.444..503N} {444, 503}

\bibitem[\protect\citeauthoryear{{Paciga} et~al.,}{{Paciga}
  et~al.}{2011}]{Paciga2011}
{Paciga} G.,  et~al., 2011, \mn@doi [\mnras]
  {10.1111/j.1365-2966.2011.18208.x}, \href
  {http://adsabs.harvard.edu/abs/2011MNRAS.413.1174P} {413, 1174}

\bibitem[\protect\citeauthoryear{{Paranjape} \& {Choudhury}}{{Paranjape} \&
  {Choudhury}}{2014}]{Paranjape2014}
{Paranjape} A.,  {Choudhury} T.~R.,  2014, \mn@doi [\mnras]
  {10.1093/mnras/stu911}, \href
  {http://adsabs.harvard.edu/abs/2014MNRAS.442.1470P} {442, 1470}

\bibitem[\protect\citeauthoryear{{Paranjape}, {Choudhury}  \&
  {Padmanabhan}}{{Paranjape} et~al.}{2016}]{Paranjape2016}
{Paranjape} A.,  {Choudhury} T.~R.,   {Padmanabhan} H.,  2016, \mn@doi [\mnras]
  {10.1093/mnras/stw1060}, \href
  {http://adsabs.harvard.edu/abs/2016MNRAS.460.1801P} {460, 1801}

\bibitem[\protect\citeauthoryear{{Park}, {Mesinger}, {Greig}  \&
  {Gillet}}{{Park} et~al.}{2019}]{Park2019}
{Park} J.,  {Mesinger} A.,  {Greig} B.,   {Gillet} N.,  2019, \mn@doi [\mnras]
  {10.1093/mnras/stz032}, \href
  {https://ui.adsabs.harvard.edu/abs/2019MNRAS.484..933P} {484, 933}

\bibitem[\protect\citeauthoryear{{Parsons}, {Pober}, {McQuinn}, {Jacobs}  \&
  {Aguirre}}{{Parsons} et~al.}{2012}]{Parsons2012}
{Parsons} A.,  {Pober} J.,  {McQuinn} M.,  {Jacobs} D.,   {Aguirre} J.,  2012,
  \mn@doi [\apj] {10.1088/0004-637X/753/1/81}, \href
  {http://adsabs.harvard.edu/abs/2012ApJ...753...81P} {753, 81}

\bibitem[\protect\citeauthoryear{{Pawlik} \& {Schaye}}{{Pawlik} \&
  {Schaye}}{2008}]{Pawlik2008}
{Pawlik} A.~H.,  {Schaye} J.,  2008, \mn@doi [\mnras]
  {10.1111/j.1365-2966.2008.13601.x}, \href
  {http://adsabs.harvard.edu/abs/2008MNRAS.389..651P} {389, 651}

\bibitem[\protect\citeauthoryear{{Petkova} \& {Springel}}{{Petkova} \&
  {Springel}}{2009}]{Petkova2009}
{Petkova} M.,  {Springel} V.,  2009, \mn@doi [\mnras]
  {10.1111/j.1365-2966.2009.14843.x}, \href
  {http://adsabs.harvard.edu/abs/2009MNRAS.396.1383P} {396, 1383}

\bibitem[\protect\citeauthoryear{{Press} \& {Schechter}}{{Press} \&
  {Schechter}}{1974}]{Press1974}
{Press} W.~H.,  {Schechter} P.,  1974, \mn@doi [\apj] {10.1086/152650}, \href
  {http://adsabs.harvard.edu/abs/1974ApJ...187..425P} {187, 425}

\bibitem[\protect\citeauthoryear{{Rai{\v c}evi{\'c}} \& {Theuns}}{{Rai{\v
  c}evi{\'c}} \& {Theuns}}{2011}]{Raicevic2011}
{Rai{\v c}evi{\'c}} M.,  {Theuns} T.,  2011, \mn@doi [\mnras]
  {10.1111/j.1745-3933.2010.00993.x}, \href
  {http://adsabs.harvard.edu/abs/2011MNRAS.412L..16R} {412, L16}

\bibitem[\protect\citeauthoryear{{Razoumov}, {Norman}, {Abel}  \&
  {Scott}}{{Razoumov} et~al.}{2002}]{Razoumov2002}
{Razoumov} A.~O.,  {Norman} M.~L.,  {Abel} T.,   {Scott} D.,  2002, \mn@doi
  [\apj] {10.1086/340451}, \href
  {http://adsabs.harvard.edu/abs/2002ApJ...572..695R} {572, 695}

\bibitem[\protect\citeauthoryear{{Santos}, {Ferramacho}, {Silva}, {Amblard}  \&
  {Cooray}}{{Santos} et~al.}{2010}]{Santos2010}
{Santos} M.~G.,  {Ferramacho} L.,  {Silva} M.~B.,  {Amblard} A.,   {Cooray} A.,
   2010, \mn@doi [\mnras] {10.1111/j.1365-2966.2010.16898.x}, \href
  {http://adsabs.harvard.edu/abs/2010MNRAS.406.2421S} {406, 2421}

\bibitem[\protect\citeauthoryear{{Semelin}, {Combes}  \& {Baek}}{{Semelin}
  et~al.}{2007}]{Semelin2007}
{Semelin} B.,  {Combes} F.,   {Baek} S.,  2007, \mn@doi [\aap]
  {10.1051/0004-6361:20077965}, \href
  {http://adsabs.harvard.edu/abs/2007A%26A...474..365S} {474, 365}

\bibitem[\protect\citeauthoryear{{Sobacchi} \& {Mesinger}}{{Sobacchi} \&
  {Mesinger}}{2014}]{Sobacchi2014}
{Sobacchi} E.,  {Mesinger} A.,  2014, \mn@doi [\mnras] {10.1093/mnras/stu377},
  \href {http://adsabs.harvard.edu/abs/2014MNRAS.440.1662S} {440, 1662}

\bibitem[\protect\citeauthoryear{{Sokasian}, {Abel}  \& {Hernquist}}{{Sokasian}
  et~al.}{2001}]{Sokasian2001}
{Sokasian} A.,  {Abel} T.,   {Hernquist} L.~E.,  2001, \mn@doi [\na]
  {10.1016/S1384-1076(01)00065-3}, \href
  {http://adsabs.harvard.edu/abs/2001NewA....6..359S} {6, 359}

\bibitem[\protect\citeauthoryear{{Stark}}{{Stark}}{2016}]{Stark2016}
{Stark} D.~P.,  2016, \mn@doi [\araa] {10.1146/annurev-astro-081915-023417},
  \href {http://adsabs.harvard.edu/abs/2016ARA%26A..54..761S} {54, 761}

\bibitem[\protect\citeauthoryear{{Thomas} et~al.,}{{Thomas}
  et~al.}{2009}]{Thomas2009}
{Thomas} R.~M.,  et~al., 2009, \mn@doi [\mnras]
  {10.1111/j.1365-2966.2008.14206.x}, \href
  {http://adsabs.harvard.edu/abs/2009MNRAS.393...32T} {393, 32}

\bibitem[\protect\citeauthoryear{{Tingay} et~al.,}{{Tingay}
  et~al.}{2013}]{Tingay2013}
{Tingay} S.~J.,  et~al., 2013, \mn@doi [\pasa] {10.1017/pasa.2012.007}, \href
  {http://adsabs.harvard.edu/abs/2013PASA...30....7T} {30, e007}

\bibitem[\protect\citeauthoryear{{Trac} \& {Cen}}{{Trac} \&
  {Cen}}{2007}]{Trac2007}
{Trac} H.,  {Cen} R.,  2007, \mn@doi [\apj] {10.1086/522566}, \href
  {http://adsabs.harvard.edu/abs/2007ApJ...671....1T} {671, 1}

\bibitem[\protect\citeauthoryear{{Trac} \& {Gnedin}}{{Trac} \&
  {Gnedin}}{2011}]{Trac2011}
{Trac} H.~Y.,  {Gnedin} N.~Y.,  2011, \mn@doi [Advanced Science Letters]
  {10.1166/asl.2011.1214}, \href
  {http://adsabs.harvard.edu/abs/2011ASL.....4..228T} {4, 228}

\bibitem[\protect\citeauthoryear{{Zahn}, {Lidz}, {McQuinn}, {Dutta},
  {Hernquist}, {Zaldarriaga}  \& {Furlanetto}}{{Zahn} et~al.}{2007}]{Zahn2007}
{Zahn} O.,  {Lidz} A.,  {McQuinn} M.,  {Dutta} S.,  {Hernquist} L.,
  {Zaldarriaga} M.,   {Furlanetto} S.~R.,  2007, \mn@doi [\apj]
  {10.1086/509597}, \href {http://adsabs.harvard.edu/abs/2007ApJ...654...12Z}
  {654, 12}

\bibitem[\protect\citeauthoryear{{Zahn}, {Mesinger}, {McQuinn}, {Trac}, {Cen}
  \& {Hernquist}}{{Zahn} et~al.}{2011}]{Zahn2011}
{Zahn} O.,  {Mesinger} A.,  {McQuinn} M.,  {Trac} H.,  {Cen} R.,   {Hernquist}
  L.~E.,  2011, \mn@doi [\mnras] {10.1111/j.1365-2966.2011.18439.x}, \href
  {http://adsabs.harvard.edu/abs/2011MNRAS.414..727Z} {414, 727}

\bibitem[\protect\citeauthoryear{{Zel'dovich}}{{Zel'dovich}}{1970}]{Zeldovich1970}
{Zel'dovich} Y.~B.,  1970, \aap, \href
  {http://adsabs.harvard.edu/abs/1970A%26A.....5...84Z} {5, 84}

\bibitem[\protect\citeauthoryear{{van Haarlem} et~al.,}{{van Haarlem}
  et~al.}{2013}]{vanHaarlem2013}
{van Haarlem} M.~P.,  et~al., 2013, \mn@doi [\aap]
  {10.1051/0004-6361/201220873}, \href
  {http://adsabs.harvard.edu/abs/2013A%26A...556A...2V} {556, A2}

\makeatother
\end{thebibliography}

\bsp	
\label{lastpage}
\end{document}